\documentclass[mathleft,fleqn,%
]{an}
\usepackage{graphicx}
\usepackage[varg]{txfonts}
\usepackage{natbib}
\bibpunct{(}{)}{;}{a}{}{,}
\begin{document}

\Pagespan{1}{}
\Yearpublication{2014}%
\Yearsubmission{2014}%
\Month{0}%
\Volume{999}%
\Issue{0}%
\DOI{asna.201400000}%

\title{Galaxy overdensities around 
  3C radio galaxies and
  quasars at 1\,$<$\,$z$\,$<$\,2.5 revealed by $\it Spitzer$
  3.6\,/\,4.5\,$\mu$m and Pan-STARRS
}

\author{
  Z. Ghaffari
  \inst{1}\fnmsep\thanks{Corresponding author:{ghaffari@astro.rub.de}}
  \and
  Ch. Westhues
  \inst{1}
  \and
  M. Haas
  \inst{1}
  \and
  R. Chini
  \inst{1,2}
  \and
  S. P. Willner
  \inst{3}
  \and
  M.L.N. Ashby
  \inst{3}
  \and
  B. J. Wilkes
  \inst{3}
}

\titlerunning{High redshift radio galaxies}
\authorrunning{Z. Ghaffari et al.}
\institute{
  Astronomisches Institut, Ruhr--Universit\"at Bochum,
  Universit\"atsstra{\ss}e 150, 44801 Bochum, Germany
  \and
  Instituto de Astronomia, Universidad Cat\'{o}lica del
  Norte, Avenida Angamos 0610, Casilla
  1280 Antofagasta, Chile
  \and
  Harvard-Smithsonian Center for Astrophysics, 60 Garden St.,
  Cambridge, MA 02138, USA
}

\received{XXXX}
\accepted{XXXX}
\publonline{XXXX}


\abstract{%
Luminous radio sources are thought to lie in galaxy clusters or 
proto-clusters. 
The complete sample of 64 high-redshift 3C sources at $1<z<2.5$ 
has been mapped with the
{\it Spitzer Space Telescope}. 
The IRAC 3.6\,$\mu$m and 4.5\,$\mu$m 5-$\sigma$ detection limit of
4\,$\mu$Jy (22.4 AB mag) 
 allows us to search 
for the 
brightest candidate cluster member galaxies associated with the 3C sources.
To remove the 
contamination of foreground 
stars and galaxies along the lines of sight toward the
3C sources 
we apply color cuts: 
removed sources satisfy 
either the IRAC1/2 cut 
$[3.6] - [4.5] < -0.3$ or the Pan-STARRS/IRAC cut 
$i - [4.5] < 0.5$ if 
 detected by Pan-STARRS.
For both selection methods, 
about half of the 3C radio sources show significant overdensities
($>3\,\sigma$)
within 30$\arcsec$ ($\sim$250\,kpc) projected distance 
from the radio source compared to the surrounding galaxy densities
measured in the 50$\arcsec$--120$\arcsec$ annulus.   
The Pan-STARRS/IRAC cut reveals higher average overdensities than
the IRAC1/2 cut  and also a decline of overdensities at $z>1.4$.
To infer the nature of the cluster members, we rerun the analysis
using a stronger IRAC1/2 cut $[3.6] - [4.5] < -0.1$ which removes
$1<z<1.4$ passive ellipticals but not star-forming galaxies. For the strong cut, the
overdensities, on average, completely disappear at $1<z<1.4$. 
We therefore suggest that the 4.5\,$\mu$m detected cluster member
galaxies are mainly passive ellipticals.
}
\maketitle

\section{Introduction} \label{sec:intro}

High-redshift ($z>1$) radio galaxies (RGs) belong to the most massive host
galaxies in the universe 
(e.g., \citet{Seymour07} and references therein)
and are 
consequently suspected to reside in galaxy proto-clusters or
clusters \citep{Miley08}. 
To confirm this idea, however, it is necessary to 
identify possible companion galaxies at
the redshifts of the RGs, in the presence of contamination 
from numerous stars 
and foreground galaxies. 
Background galaxies unless gravitationally lensed become 
fainter on average with increasing distance
and therefore are expected to provide a relatively minor contamination. 
Previous studies have found galaxy overdensities around distant active
galactic nuclei (AGN), but
the results are far from definitive. 

Multi-band imaging and photometric redshift estimation
are efficient ways to find candidate cluster member
galaxies.
In order to yield reliable outcomes, photo-$z$ techniques generally 
require five or more photometric measurements, and ideally these
should encompass  
a strong and ubiquitous
feature in the spectral energy distribution (SED), e.g., the
4000\,\AA\ Balmer break. 
If fewer SED data points are available, however, broadband color cuts are a widely used 
alternative for selecting high redshift sources as done by, e.g., \citet{Huang04}.

The SEDs of nearly all galaxies show a peak at 1--2\,$\mu$m in the rest frame,
making near-infrared (NIR) observations very useful for photo-$z$
estimates; for a detailed discussion we refer the reader to \citet{Galametz12}. 
Pioneering $K$ and $J$ band imaging of high-$z$ 3C sources indicated
a mean galaxy 
overdensity but with a large field-to-field scatter \citep{Best00, Best03}. These 
results were obtained with relatively small area coverage, limited to 50$\arcsec$ field of view (FoV), 
only about the expected cluster size at $z=1$, thus making the estimation 
of the nearby foreground (and background) uncertain.
Furthermore, at redshifts above 1, the SED peak shifts out of the $K$ band,  
and observations at longer wavelengths are advantageous for source detection. 

Such observations and studies have been performed 
for the sample of 72 RGs at $1<z<5.2$, hereafter referred to as the
{\it Spitzer}-HzRG 
sample (\citealt{Seymour07}, \citealt{Galametz12}) and an
extension to 387 radio-loud AGNs at $1<z<3.2$, hereafter referred to as the CARLA
sample \citep{Wylezalek13}.
For these  
the environment has been studied with observations at 3.6
and 4.5\,$\mu$m using the IRAC instrument 
\citep{Fazio04} onboard 
the {\it Spitzer Space Telescope} \citep{Werner04}.
About half of the RGs and quasars in these samples show an overdensity of 
{\it Spitzer}/IRAC-detected galaxies, both compared to blank fields
and in terms of a 
surface density rising toward the position of the AGN 
\citep{Galametz12, Wylezalek13, Hatch14}. 
However, the other half shows only little evidence for associated clusters.
The studies successfully used an IRAC color cut 
$[3.6] - [4.5] > -0.1$ (AB mag), following 
\citet{Huang04}
to identify $z>1.3$ candidate cluster member galaxies.

\citet{Falder10} employed an efficient strategy based on
IRAC 3.6\,$\mu$m one-filter-only observations of 173 AGN at $z \sim 1$
to reveal 
an excess of
massive galaxies within 300\,kpc of their targets compared to the
field.  The excess is more pronounced for radio-loud AGN.
Notably, \citet{Falder10} did not use any color 
information to reject foreground sources.

\begin{table*}
\renewcommand{\thetable}{\arabic{table}}
\centering
\caption{3C sample at $1 < z < 2.5$ from \citet{Spinrad85}.
} 
\label{tab_sample}
\begin{tabular}{lcccc|lcccc}
\hline
\hline
Name & RA (J2000) & Dec (J2000)& Type$^1$ & Redshift          & Name & RA (J2000) & Dec (J2000)& Type$^1$ & Redshift \\
\hline
 3C\,002      & 00h06m22.6s  & -00d04m25s  &   Q  &   1.037   &   3C\,250      & 11h08m52.1s  & +25d00m55s  &   G  &   1.260  \\
 3C\,009      & 00h20m25.3s  & +15d40m53s  &   Q  &   2.009   &   3C\,252      & 11h11m33.1s  & +35d40m42s  &   G  &   1.100  \\
 3C\,013      & 00h34m14.5s  & +39d24m17s  &   G  &   1.351   &   3C\,255      & 11h19m25.2s  & -03d02m52s  &   Q  &   1.355  \\
 3C\,014      & 00h36m06.5s  & +18d37m59s  &   Q  &   1.469   &   3C\,256      & 11h20m43.0s  & +23d27m55s  &   G  &   1.819  \\
 3C\,036      & 01h17m59.5s  & +45d36m22s  &   G  &   1.301   &   3C\,257      & 11h23m09.2s  & +05d30m19s  &   G  &   2.474  \\
 3C\,043      & 01h29m59.8s  & +23d38m20s  &   Q  &   1.459   &   3C\,266      & 11h45m43.4s  & +49d46m08s  &   G  &   1.275  \\
 3C\,065      & 02h23m43.2s  & +40d00m52s  &   G  &   1.176   &   3C\,267      & 11h49m56.5s  & +12d47m19s  &   G  &   1.140  \\
 3C\,068.1    & 02h32m28.9s  & +34d23m47s  &   Q  &   1.238   &   3C\,268.4    & 12h09m13.6s  & +43d39m21s  &   Q  &   1.397  \\
 3C\,068.2    & 02h34m23.8s  & +31d34m17s  &   G  &   1.575   &   3C\,270.1    & 12h20m33.9s  & +33d43m12s  &   Q  &   1.532  \\
 3C\,119      & 04h32m36.5s  & +41d38m28s  &   G  &   1.023   &   3C\,280.1    & 13h00m33.3s  & +40d09m08s  &   Q  &   1.671  \\
 3C\,124      & 04h41m59.1s  & +01d21m02s  &   G  &   1.083   &   3C\,287      & 13h30m37.7s  & +25d09m11s  &   Q  &   1.055  \\
 3C\,173      & 07h02m17.6s  & +37d57m20s  &   G  &   1.035   &   3C\,294      & 14h06m44.0s  & +34d11m25s  &   G  &   1.779  \\
 3C\,181      & 07h28m10.3s  & +14d37m36s  &   Q  &   1.382   &   3C\,297      & 14h17m24.0s  & -04d00m48s  &   G  &   1.406  \\
 3C\,186      & 07h44m17.4s  & +37d53m17s  &   Q  &   1.067   &   3C\,298      & 14h19m08.2s  & +06d28m35s  &   Q  &   1.437  \\
 3C\,190      & 08h01m33.5s  & +14d14m42s  &   Q  &   1.195   &   3C\,300.1    & 14h28m31.3s  & -01d24m08s  &   G  &   1.159  \\
 3C\,191      & 08h04m47.9s  & +10d15m23s  &   Q  &   1.956   &   3C\,305.1    & 14h47m09.5s  & +76d56m22s  &   G  &   1.132  \\
 3C\,194      & 08h10m03.6s  & +42d28m04s  &   G  &   1.184   &   3C\,318      & 15h20m05.4s  & +20d16m06s  &   G  &   1.574  \\
 3C\,204      & 08h37m44.9s  & +65d13m35s  &   Q  &   1.112   &   3C\,322      & 15h35m01.2s  & +55d36m53s  &   G  &   1.681  \\
 3C\,205      & 08h39m06.4s  & +57d54m17s  &   Q  &   1.534   &   3C\,325      & 15h49m58.6s  & +62d41m21s  &   G  &   1.135  \\
 3C\,208.0    & 08h53m08.8s  & +13d52m55s  &   Q  &   1.110   &   3C\,324      & 15h49m48.9s  & +21d25m38s  &   G  &   1.206  \\
 3C\,208.1    & 08h54m39.3s  & +14d05m53s  &   G  &   1.020   &   3C\,326.1    & 15h56m10.1s  & +20d04m20s  &   G  &   1.825  \\
 3C\,210      & 08h58m09.9s  & +27d50m52s  &   G  &   1.169   &   3C\,356      & 17h24m19.0s  & +50d57m40s  &   G  &   1.079  \\
 3C\,212      & 08h58m41.5s  & +14d09m44s  &   Q  &   1.048   &   3C\,368      & 18h05m06.3s  & +11d01m33s  &   G  &   1.131  \\
 3C\,220.2    & 09h30m33.5s  & +36d01m24s  &   Q  &   1.157   &   3C\,418      & 20h38m37.0s  & +51d19m13s  &   Q  &   1.686  \\
 3C\,222      & 09h36m32.0s  & +04d22m10s  &   G  &   1.339   &   3C\,432      & 21h22m46.2s  & +17d04m38s  &   Q  &   1.785  \\
 3C\,225A     & 09h42m08.5s  & +13d51m54s  &   G  &   1.565   &   3C\,437      & 21h47m25.1s  & +15d20m37s  &   G  &   1.480  \\
 3C\,230      & 09h51m58.8s  & -00d01m27s  &   G  &   1.487   &   3C\,454.0    & 22h51m34.7s  & +18d48m40s  &   Q  &   1.757  \\
 3C\,238      & 10h11m00.4s  & +06d24m40s  &   G  &   1.405   &   3C\,454.1    & 22h50m32.9s  & +71d29m19s  &   G  &   1.841  \\
 3C\,239      & 10h11m45.4s  & +46d28m20s  &   G  &   1.781   &   3C\,469.1    & 23h55m23.3s  & +79d55m20s  &   G  &   1.336  \\
 3C\,241      & 10h21m54.5s  & +21d59m30s  &   G  &   1.617   &   3C\,470      & 23h58m35.3s  & +44d04m39s  &   G  &   1.653  \\
 3C\,245      & 10h42m44.6s  & +12d03m31s  &   Q  &   1.028   &   4C\,13.66    & 18h01m39.0s  & +13d51m23s  &   G  &   1.450  \\
 3C\,249      & 11h02m03.8s  & -01d16m17s  &   Q  &   1.554   &   4C\,16.49    & 17h34m42.6s  & +16d00m31s  &   Q  &   1.880  \\
\hline
\end{tabular}
~\\
$~^1$ Type denotes quasar (Q) or radio galaxy (G), depending on
whether or not broad emission lines have been identified in their spectra.
\end{table*}

A more elaborate study by \citet{Haas09} of 3C\,270.1 at $z=1.54$
involves {\it Spitzer}/IRAC 
observations in combination with deep $z$ and $y$ band imaging 
($\sim$26 AB
mag) with the 6.5\,m MMT at Mt.\ Hopkins 
\citep{McLeod06, Brown08}. 
This strategy, which brackets the Balmer break at the expected cluster
redshift, is relatively insensitive to photometric errors in the IRAC
bands, which 
easily may exceed 10\%.
Fitting $zy$+IRAC photometry with two simple galaxy
templates (elliptical and ULIRG)
revealed
a central overdensity of intrinsically 
red $z \approx 1.54$ galaxies around the
radio source. 
This technique therefore shows significant promise for identifying 
candidate cluster member galaxies, 
but the need to collect the required deep
optical imaging has so far impeded a systematic investigation of this kind 
for large galaxy samples. 

\begin{figure*}
  \includegraphics[width=17cm, clip=true]{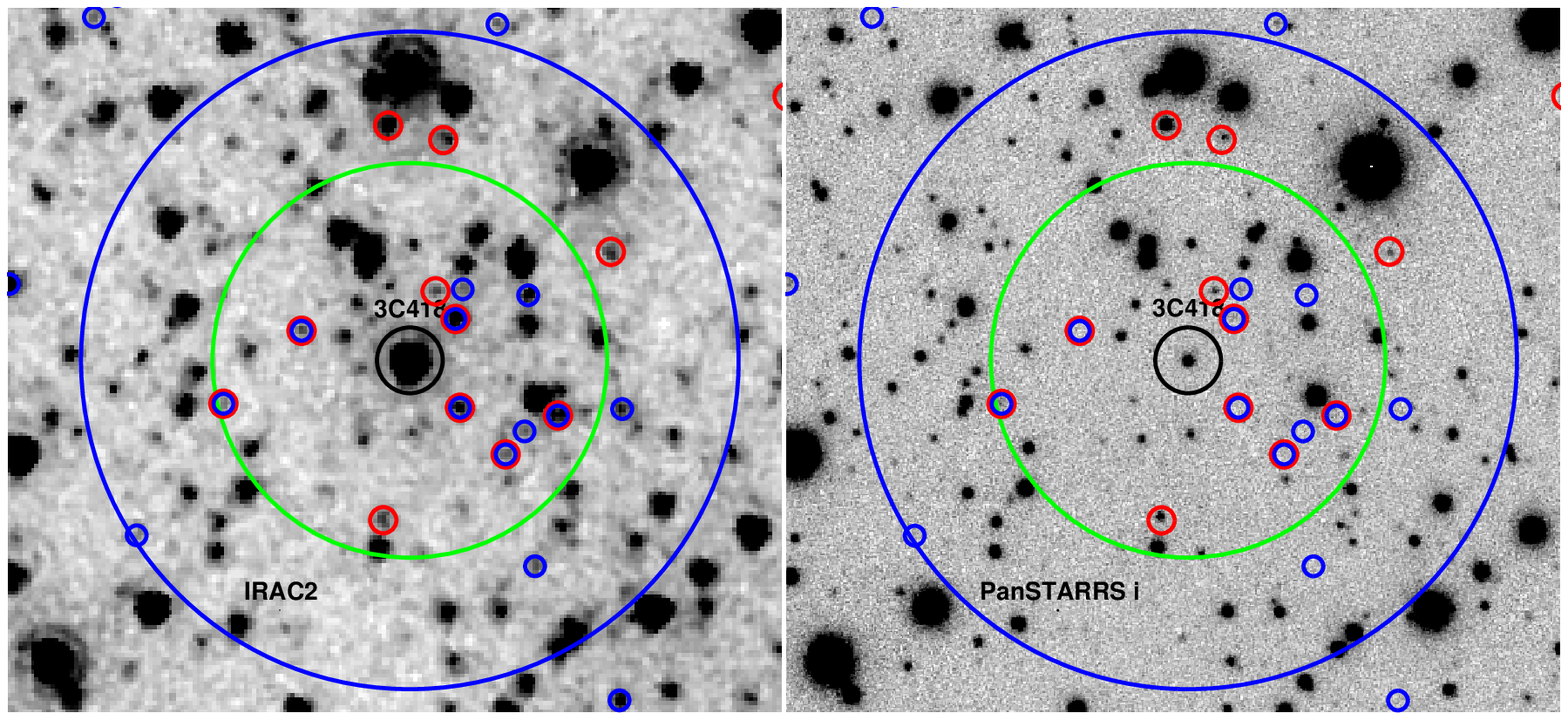}
  \includegraphics[width=17cm, clip=true]{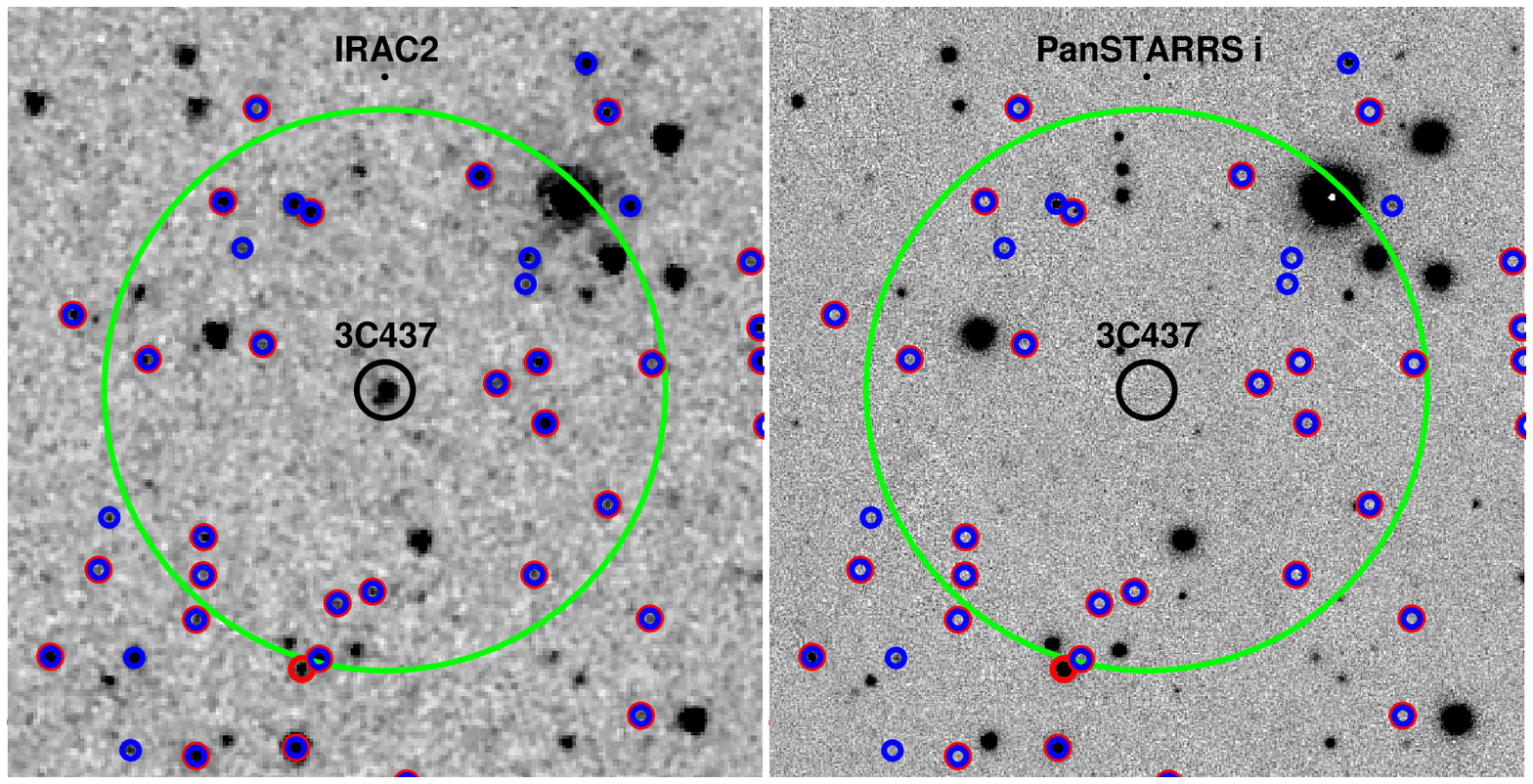}
  \caption{
   Images of the most crowded field (top: quasar 3C\,418) and a sparsely populated
   field (bottem, radio galaxy 3C\,437). 
   The left panels are IRAC2 at 4.5$\mu$m, the right panels are Pan-STARRS $i$-band.
   The black circles of 5$\arcsec$ radius mark the 3C sources.
   The large circles (green/blue) have a radius of 30$\arcsec$ and 50$\arcsec$, respectively.
   The small circles mark the candidate cluster member galaxies 
   found in Section~\ref{sec:galaxy_overdensities} with the Pan-STARRS cut
   (blue, 1$\farcs$5) and the IRAC cut (red, 2$\arcsec$ radius).
  }
  \label{fig_images}
\end{figure*}

The 3C sample contains 64 radio galaxies and quasars at $1<z<2.5$ 
\citep{Spinrad85}. They all have been observed with {\it Spitzer}/IRAC 
($3.6-8 \mu$m) and MIPS (24\,$\mu$m) with a nearly 100\% detection
rate \citep{Podigachoski15}. 
While six 3C sources are contained in the {\it Spitzer}-HzRG sample
(mentioned above),  
most of the high-$z$ 3C environments have not been investigated.
In addition, the high-$z$ 3C sample has been mapped with the {\it Herschel
Space Observatory} with a detection rate of over 50\%  \citep{Podigachoski15}, in contrast to
the $\sim$20\% {\it Herschel} detection rate of the {\it Spitzer}-HzRG sample
\citep{Drouart14}. 
The high {\it Herschel} detection rate makes the 3C sample particularly 
interesting for an environment
study over the whole near-, mid-, {\it and} far-infrared wavelength range.
Such a study will allow one also to explore a possible connection between the
environment and the black hole and star forming activity of the
radio-loud AGNs and their host galaxies.

The advent of the Pan-STARRS $grizy$ data release 
\citep{Chambers16, Magnier16} offers a unique opportunity
for a joint {\it Spitzer} $-$ Pan-STARRS study of the high-$z$ 3C environments.
While the depth of Pan-STARRS (with a completeness limit of about $i =
21.2$\,AB mag, see Section~\ref{sec:data} below)  
might be too shallow to
detect candidate cluster member galaxies, the Pan-STARRS data
nonetheless help to reject many or even most of the foreground sources.

We here do not distinguish between
clusters, proto-clusters and galaxy groups, rather simply search for galaxy overdensities.
We examine the environment of 3C radio sources as seen in the near- and
mid-infrared bands provided by {\it Spitzer}/IRAC+MIPS, supplemented in the
visible by Pan-STARRS (PSO).
This paper is organized as follows.  Section~\ref{sec:data} 
describes the sample and data, including 
the matched {\it Spitzer}$-$Pan-STARRS
source catalog and the correction for the foreground galactic extinction.
Section~\ref{sec:sed_examples} shows example SEDs.
Color cuts for redshifted galaxy templates are described in
Section~\ref{sec:color_cuts}. Section~\ref{sec:galaxy_overdensities}
reports of two approaches to find galaxy overdensities in the high-$z$ 
3C environments. 
A summary is presented in Section~\ref{sec:summary}.
Throughout this work we adopt a standard $\Lambda$CDM cosmology
($H_\circ$ = 70 km s$^{-1}$ 
Mpc$^{-1}$, $\Omega_\Lambda = 0.73$, and $\Omega_m = 0.27$ \citep{Spergel07}. All
magnitudes are AB where zero mag corresponds to 3631~Jy. 

\section{Data} \label{sec:data}

Figure~\ref{fig_images} shows IRAC2 and Pan-STARRS $i$- band images of
a crowded and a sparsely populated field. Only 5\% of the 3C sources are
that much crowded like 3C\,418. About 60\% of the sample has intermediately
crowded fields, and  35\% of the 3C sources lie in sparse fields like 3C\,437. 
Figure~\ref{fig_images} illustrates that our images are rather shallow and that most
sources are well isolated, with the advantage that source confusion
and deblending plays a minor role. 
It also indicates that the PSF differences (see Table~\ref{tab_detection_limits})
between the filters do not pose a real challenge for the matching. 
This conclusion is further
supported by the fact that in the final catalog of 50000 IRAC sources only
about 15\% of the IRAC sources have a double or multiple Pan-STARRS match.

\subsection{{\it Spitzer}/IRAC+MIPS} \label{sec:data_spitzer}

The sample of 64 ~ 3C sources at spectroscopic redshifts $1<z<2.5$ 
is listed in Table.~\ref{tab_sample}. All sources were mapped
by the {\it Spitzer Space Telescope} during the cryogenic mission as part of several guaranteed time 
programs led by IRAC PI Giovanni Fazio
and described in more detail by \citet{Seymour07} and \citet{Haas08}. 
In brief, the observations include the four IRAC bands 
(3.6, 4.5, 5.8, and 8.0\,$\mu$m), yielding three fields with roughly 
4$\farcm$5 $\times$ 4$\farcm$5 field of view (FoV).  The central
field was covered with all four filters, and the two flanking fields 
were each observed in two filters, with
one about 6\,$\farcm$5 to the "left" in 3.6 and 5.8\,$\mu$m and the 
other to the "right" in 4.5 and 8.0\,$\mu$m.
In addition, 24\,$\mu$m 
imaging of the central field covering about
5$\arcmin$ $\times$ 5$\arcmin$ FoV was also obtained using the MIPS instrument \citep{Rieke04}. 
The total exposure times were $4 \times 30$\,s for each IRAC band and
$10 \times 10$\,s for the MIPS images. Eight 3C sources contained in the {\it
  Spitzer}-HzRG program  
(3C\,065, 3C\,239, 3C\,257, 3C\,266, 3C\,294, 3C\,356, 3C\,368, 3C\,470) 
were observed at similar depths but covering a larger area. 
Photometry of the 3C sources themselves has already been reported 
by \citet{Seymour07} for 
eight sources and by \citet{Haas08} for the remaining 56 sources. 
\citet{Podigachoski15} published the
{\it Spitzer} (and {\it Herschel}) photometry.

For the present study, we used the {\it Spitzer} Enhanced Imaging Products
(SEIP)\footnote{http://irsa.ipac.caltech.edu/data/SPITZER/Enhanced/SEIP/
overview.html}
to identify all sources in the IRAC+MIPS mosaics and carry out photometry.
In each band, mosaic maps were created for sky areas covered by at least
four dithered images. 
The pixel size of the provided IRAC maps is 0$\farcs$6 (half the
native IRAC pixel size) and for the 
MIPS 24\,$\mu$m it is 2$\farcs$45 (native MIPS\,24 pixel size). 
We used the SEIP average maps (with artifacts having been removed by
SEIP algorithms) because the
background of the average maps is slightly smoother than that of the
median maps.  

\subsection{{\it Spitzer} Source Extraction and Photometry}
\label{sec:data_spitzer_extraction}

SEIP provides source catalogs for 10-$\sigma$
detections with photometry in 3$\farcs$8 and 5$\farcs$8 diameter apertures, 
but we found by visual inspection that numerous visible sources are 
missing from the SEIP catalogs.
We therefore performed new source identification 
and photometry using SExtractor\footnote{http://www.astromatic.net/software/sextractor}  
\citep{Bertin96} in a manner similar to that 
described by \citet{Baronchelli16} for the {\it Spitzer}/IRAC+MIPS 
Extragalactic Survey (SIMES) but with several small modifications. 
We used a spacing of 128 $\times$ 128 pixels
filtered with a 
 3-pixel top-hat kernel to model the backgrounds. This was necessary to 
account for extended sources ($\sim$5\%) present in our maps. 
We used a pixel detection threshold of 2$\sigma$ and a minimum number of six
connected pixels above the background noise.  
We computed AUTO fluxes as done by Baronchelli et al.\footnote{AUTO
  fluxes 
  are estimates of the total flux of a
  source in an 
  elliptical aperture of semi-major axis ($a$) proportional to the Kron
  radius  $R_K$ of the object \citep{Kron80}. Setting 
  the SExtractor parameter KronFact = 2.5, hence $a = 2.5 R_K$ means -- by
  definition of the Kron
  radius -- that the AUTO flux contains more
  than 90\% of the total galaxy flux. 
  For apertures with $a < 3.5$ pixels (2$\farcs$1), the AUTO flux was
  computed within a circular aperture.
  AUTO fluxes account for the extent of each source, the
  elliptical shapes of the observed isophotes and the source's radial
  surface profile. 
  For close double sources, AUTO flux perfoms a deblending and 
  returns the Kron flux assuming
  isophotal symmetry. 
}

\begin{figure}
  \includegraphics[width=8cm, clip=true]{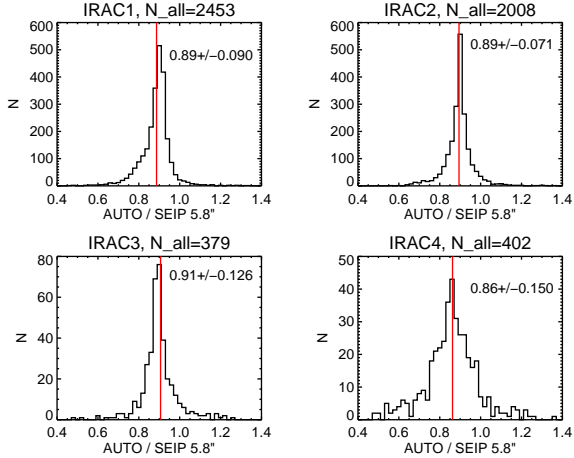}
  \caption{
   Ratio of our AUTO flux to total 5$\farcs$8 aperture flux of the
   SEIP catalog. Only compact, isolated sources enter 
   the histograms.
   The vertical red line marks the average.
   Average and 1-$\sigma$ are as labelled.
  }
  \label{fig_comparison_seip}
\end{figure}

In a first step we ran SExtractor on each IRAC image in
single mode. 
We checked the positional accuracy of sources in the
various IRAC bands in order to see how well the sources match.
The IRAC positions are generally good to
less than $0\farcs2$ which means that position-matching to
within 1$\arcsec$ is appropriate, 
as also found by \citet{Ashby13} for the Spitzer
Extended Deep Survey. 
We checked for each IRAC image and each source the
separation to the nearest neighbouring source.
It turns out that depending on the filter between 10\% and 20\%
of the sources have a neighbour 
closer than 6$\arcsec$, many with even a 4$\arcsec$ neighbour. 
This rules out the use of
apertures larger than about 4$\arcsec$ to avoid flux contamination by
the neighbour. 
On the other hand, about 5\%  of the  
sources are more extended than 4$\arcsec$, so that flux is lost when
performing photometry with less than 4$\arcsec$
diametre aperture.
Therefore all photometry given here refers to the AUTO fluxes which
return the deblended Kron flux assuming isophotal symmetry. 

In a second pair of steps, we ran SExtractor  
in dual mode using first
the 4.5\,$\mu$m images and then the 3.6\,$\mu$m images as a reference. 
Using 4.5\,$\mu$m (3.6\,$\mu$m) as reference allows us to detect the 
sources in the central
field and in the right (left) off-field, respectively.
In general, the 4.5\,$\mu$m images (and the other longer wavelength 
images) are less affected by
mux-bleeds from bright saturated stars compared to the 3.6\,$\mu$m images. 
Therefore we used as primary catalogs those based on the 4.5\,$\mu$m
images (center fields and right off-fields) and added sources in the
left off-fields from the 3.6\,$\mu$m--based 
catalogs.\footnote{While both the left and the right off-fields
  will not be used in the environment analysis 
in Section~\ref{sec:galaxy_overdensities}, they are
cataloged here as well for future use.}
To get rid of false detections in the reference frames, 
we required that a source must be detected in single mode in at
least one other filter (using a matching radius of 1$\arcsec$). 
Finally, only sources with SExtractor detection flag $\le$ 3 were
accepted; this rejects saturated sources and uncertain sources near
the map border.

To check on the uncertainties of the IRAC photometry we 
compared our AUTO fluxes with the SEIP pipeline
catalog photometry, which lists aperture corrected total fluxes
measured with circular  
apertures of 3$\farcs$8 and 5$\farcs$8 
diameter. 
We used only sources which are compact (with Kron radius smaller than
6$\arcsec$) and isolated (with a nearest neighbour more than
10$\arcsec$ apart) and required that the 3$\farcs$8 and 5$\farcs$8
photometry agrees to better than 10\%.
Figure~\ref{fig_comparison_seip} 
shows the ratio of 
AUTO flux to total flux  for the 5$\farcs$8 aperture.
For all bands the average ratio of 
AUTO flux to total flux is $\sim$0.9, 
consistent with the Kron definition of the AUTO flux.
The results are similar for the 3$\farcs$8 aperture
These validation checks document, on average, the good photometric
quality and that any remaining filter dependent PSF effects are
remarkably well corrected for. 
Individual sources, however, may deviate  
considerably (about 30\%).
This demands for caution when our data are used for applying narrow IRAC1/2
color cuts which 
may be sensitive to photometric errors larger than 10\%
(as discussed further in Section~\ref{sec:color_cuts}). 

\begin{figure*}
  \hspace{-5mm}\includegraphics[width=6cm, clip=true]{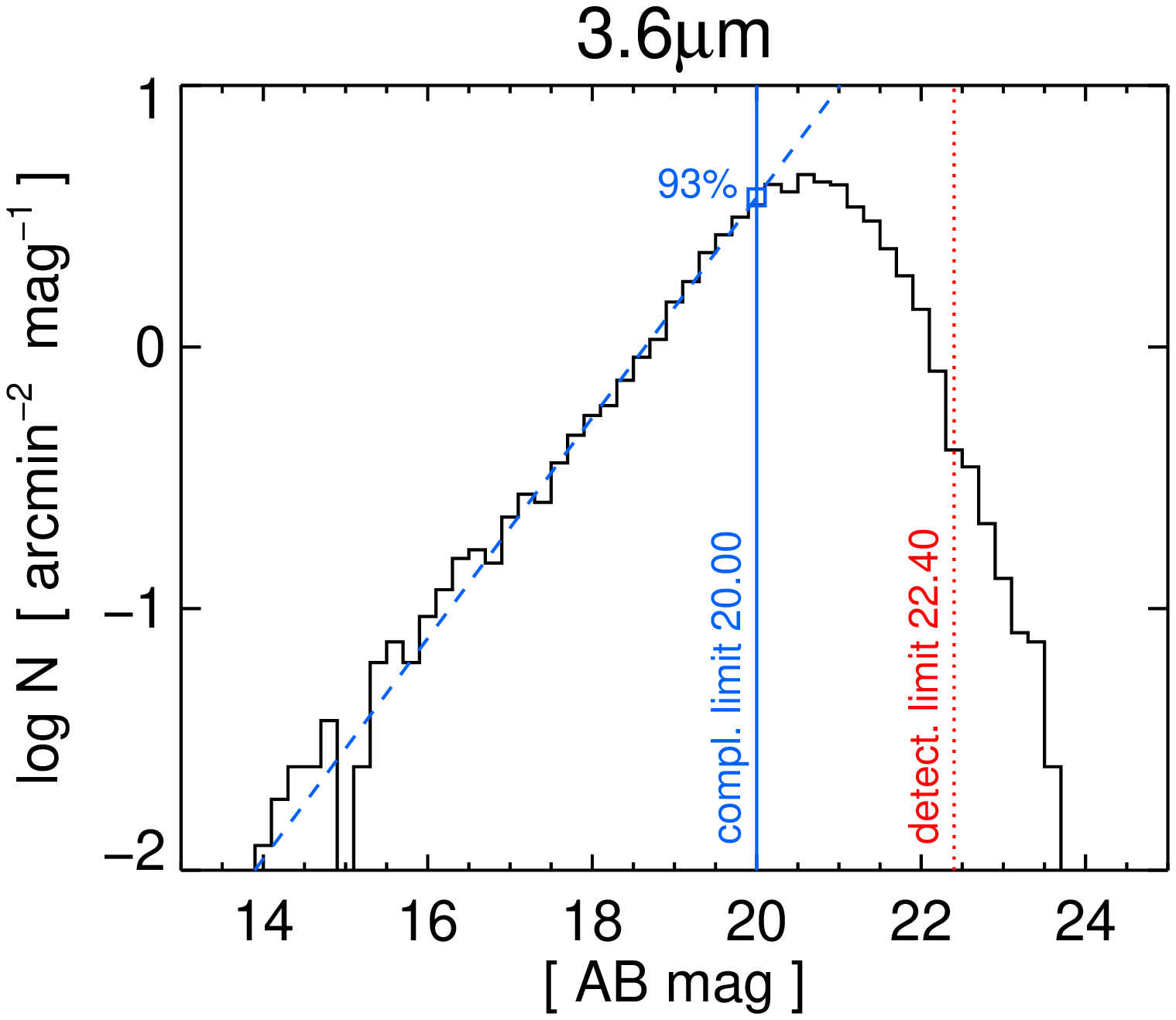}
  \hspace{-5mm}\includegraphics[width=6cm, clip=true]{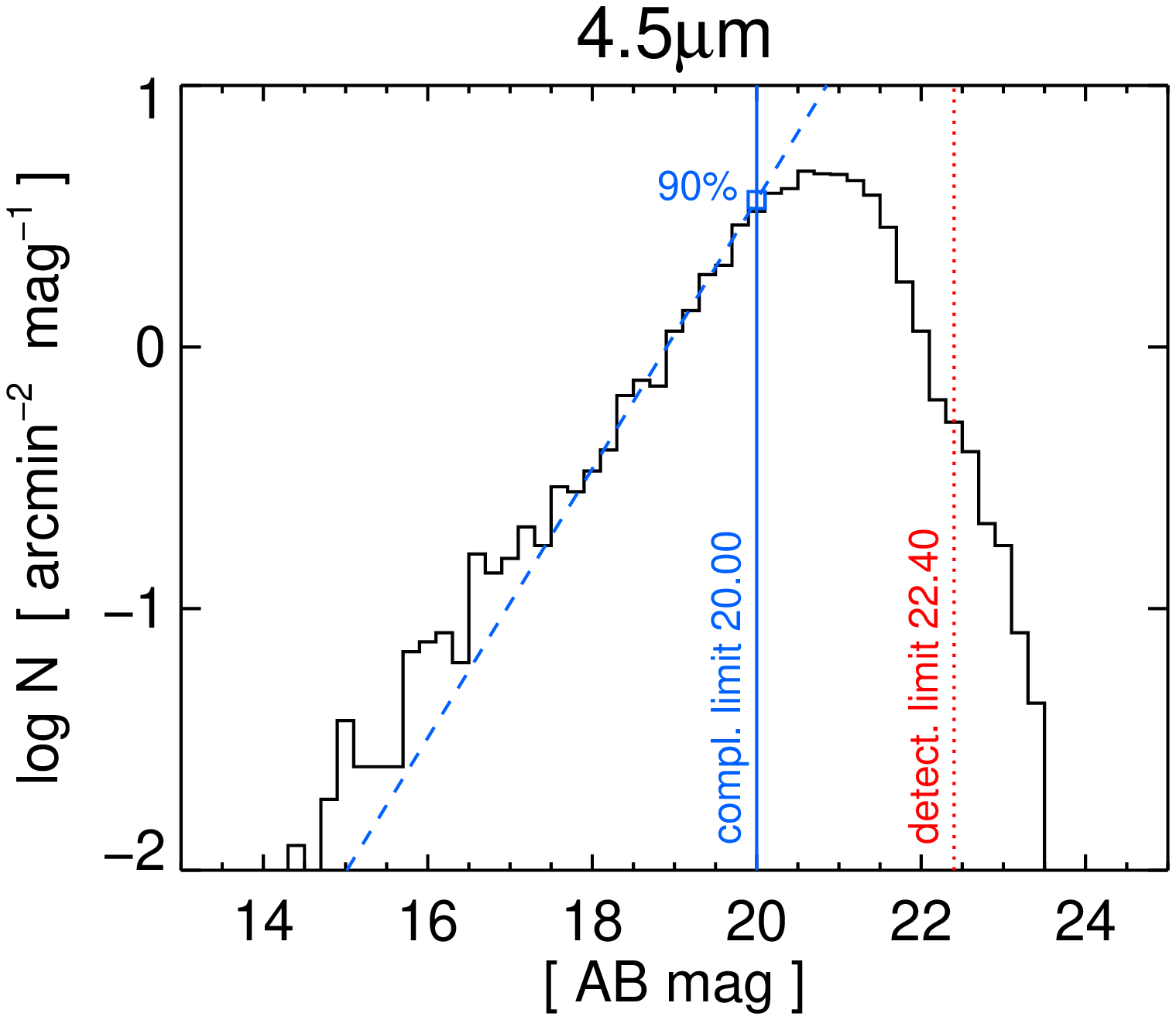}
  \hspace{-5mm}\includegraphics[width=6cm, clip=true]{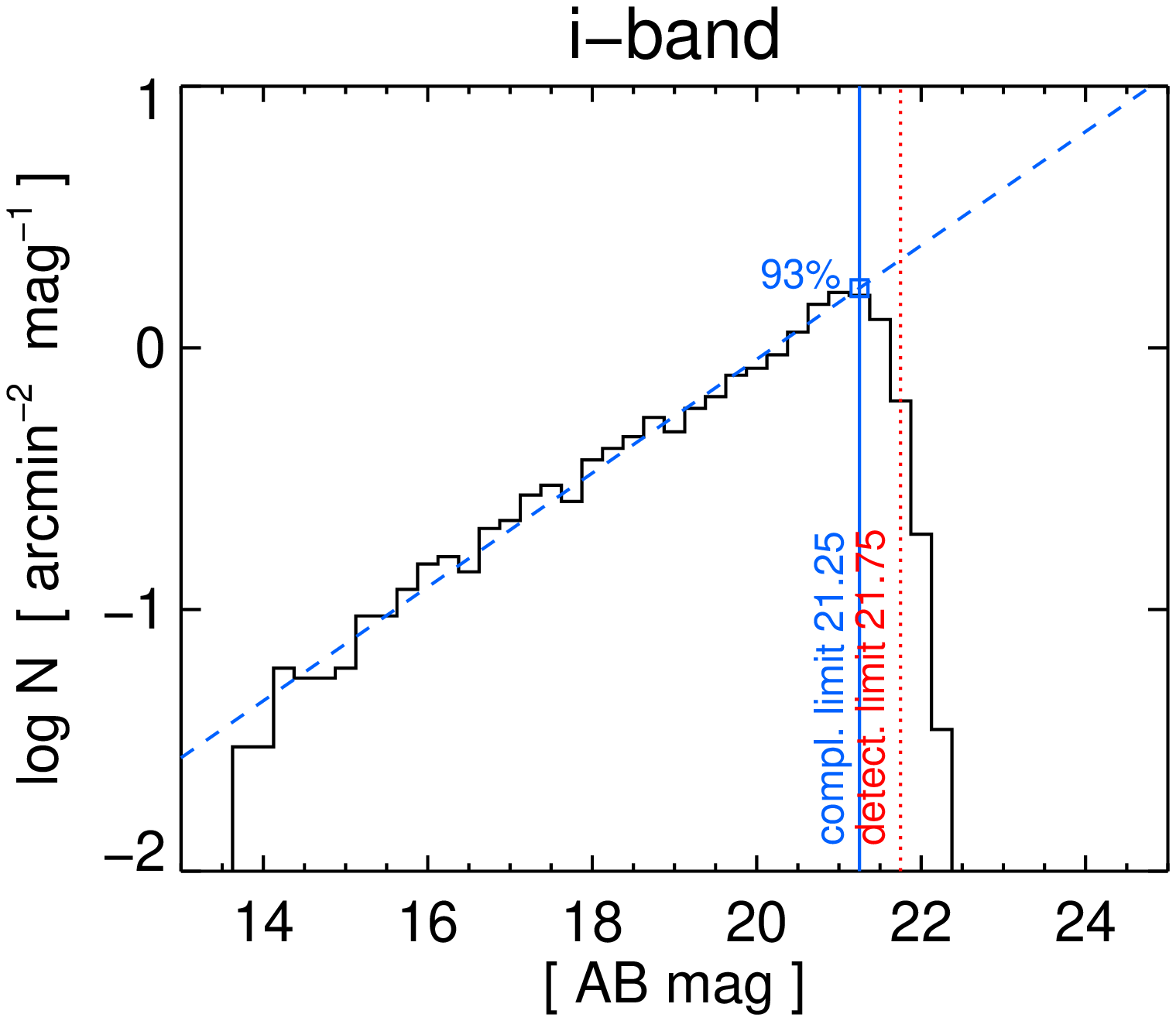}
  \caption{ Number counts within 2$\arcmin$ radius around all 3C sources in the three filters
    3.6\,$\mu$m, 4.5\,$\mu$m and $i$ used for the galaxy overdensity search.
    The detection limit (red dotted line) is the magnitude at which 
    98\% of the sources are brighter. 
    The blue dashed line shows a linear fit.
    The completeness limit (blue solid line) and 
    the completeness fraction ($\sim$90\%) are estimated relative to the
    extrapolation of the linear fit (blue square).
  }
  \label{fig_detection_limits}
\end{figure*}

For the MIPS 24\,$\mu$m images we ran SExtractor in single-image mode
only, using a global background of 32 pixels filtered with a
3 pixel top-hat kernel. To improve the detection and
photometry  and to reject uncertain sources at the map border, 
sources were accepted only if the coverage map indicated at least 6
exposures (maximum is 10 exposures). 

We estimated the sensitivity and completeness limits via
number count histograms of sources detected with at least $5-\sigma$
(Fig.~\ref{fig_detection_limits}). 
As detection limit we did not simply take the faintest detected source,
rather we allowed for 2\% faint outliers, thus 98\% of the sources are
brighter than the detection limit.
The values are listed in Table~\ref{tab_detection_limits}.
Both detection and completeness limits 
agree within 0.2 mag for
all 64 central fields.\footnote{Both detection and completeness limits
  are about 0.5--0.7\,mag brighter for the IRAC off-fields.
  This is because we required that a source must be detected in single
  mode in at least two filters, and that the available second filter
  in the off-fields (5.8 and 8$\mu$m,respectively) is less sensitive than the 
  3.6 and 4.5$\mu$m bands.
} 

\begin{table}
\renewcommand{\thetable}{\arabic{table}}
\centering
\caption{5-$\sigma$ detection limits and 
  completeness limits in AB mag for each filter as determined from 
  histograms such as shown in Figure~\ref{fig_detection_limits}. 
  Zero mag corresponds to 3631\,Jy.
  The last column lists the PSF FWHM.
  For Pan-STARRS median values from \citet{Magnier16} are given.
} 
\label{tab_detection_limits}
\setlength{\tabcolsep}{2.7pt}
\begin{tabular}{ccccc}
\hline
\hline
Filter & Wavelength & Detect. & Compl.       & PSF FWHM \\
        & [$\mu$m] &          & ($\sim$90\%) & arcsec \\
\hline
PSO $g$  & 0.481    & 22.2  & 21.7  &  $\sim$1.1     \\
PSO $r$  & 0.617    & 22.0  & 21.5  &  $\sim$1.1     \\
PSO $i$  & 0.752    & 21.7  & 21.2  &  $\sim$1.1     \\
PSO $z$  & 0.866    & 21.2  & 20.5  &  $\sim$1.1     \\
PSO $y$  & 0.962    & 20.5  & 19.7  &  $\sim$1.1     \\
IRAC 1   & 3.6      & 22.4  & 20.0  &  1.66          \\
IRAC 2   & 4.5      & 22.4  & 20.0  &  1.72          \\
IRAC 3   & 5.8      & 20.8  & 19.4  &  1.88          \\
IRAC 4   & 8.0      & 20.8  & 19.6  &  1.98          \\
MIPS 1   &24.0      & 18.5  & 17.3  &  $\sim$5.9     \\
\hline
\end{tabular}
\end{table}

\subsection{Pan-STARRS} \label{sec:data_panstarrs}

Pan-STARRS has imaged the entire northern sky in the five bands $g,r,i,z,y$
and created catalogs from 
the image data\footnote{https://panstarrs.stsci.edu/}
\citep{Chambers16, Flewelling16}.
We retrieved the Pan-STARRS images (with a pixel scale of 0$\farcs$25) and source
catalogs for all 64 high-$z$ 3C fields 
from the Mikulski Archive for Space Telescopes (MAST) at Space
Telescope Science Institute web  
page\footnote{https://archive.stsci.edu/panstarrs/}.
From the source catalogs we extracted sources within 12$\arcmin$ of
the 3C position 
and used the Kron photometry \citep{Magnier16}. 

\subsection{Cross-matched catalog} \label{sec:catalog}

We found systematic positional offsets, 
median about 0$\farcs$2, between
IRAC and Pan-STARRS sources.\footnote{The net position offset 
is consistent with the
offset between 2MASS and Pan-STARRS see \citet{Magnier16}.  
IRAC positions are tied to
2MASS.}
Therefore, for each 3C field, we performed two iterations of the
matching: in the first 
iteration we determined the median offset between IRAC and Pan-STARRS
using a matching radius of 1$\farcs$2.
In the second, we matched the offset-corrected positions using a matching radius
of 1$\arcsec$.
Among all sources finally matching within 1$\arcsec$, 96\% match
within 0$\farcs$75 and  87\% within 0$\farcs$5.
On average, about 15\% of the IRAC sources have a
double or multiple PSO counterpart.
These sources are flagged and not used in the overdensity analysis.
For matching the MIPS24 and IRAC sources we used 
a radius of 2$\arcsec$.\footnote{Increasing the 2$\arcsec$ MIPS24 --
  IRAC matching 
radius to 3$\arcsec$ did not increase the number of unique matches.}

The final {\it Spitzer} source catalog, supplemented by Pan-STARRS
photometry contains
about 50000 sources. About 20000 sources are detected with both
{\it Spitzer} and Pan-STARRS, the remaining 30000 sources only 
with {\it Spitzer}.\footnote{There are also 200,000 Pan-STARRS sources 
 not detected by {\it Spitzer}. Those are of no interest for present
 purposes.}  
About 27000 sources are located in the IRAC central fields and 
detected in both IRAC 3.6 and 4.5\,$\mu$m bands 
(on average 20 sources per square arcmin); 10000 of these 27000 
sources are detected also with Pan-STARRS.
In the IRAC off-fields the detection rate is lower because
only two IRAC filters were observed; we required sources to be
detected in both filters in order to exclude spurious sources.

Among those
sources of the present {\it Spitzer}$-$Pan-STARRS catalog 
detected at both 3.6 and 4.5\,$\mu$m, the Pan-STARRS
detection rate in the $g,r,i, z,y$-bands is 24, 33, 40, 37, 29\%, respectively. 
The $i$-band has the highest detection rate, 
and only $\sim$4\% of the $i$-band non-detections were detected in any of
the other four Pan-STARRS bands. 
Following the same procedure used for the {\it Spitzer} bands, we
determined the  
Pan-STARRS detection and completeness limits via
histograms for those sources in the {\it Spitzer}$-$Pan-STARRS catalog
(Fig.~\ref{fig_detection_limits}, Tab.~\ref{tab_detection_limits}).

We did not apply a correction for PSF differences between the
filters.
We consistently use Kron photometry for both Pan-STARRS and {\it Spitzer}
data. We expect any bias to be small compared to the photometric
uncertainty of the faint IRAC 
sources (10-30\%). 

The catalog was created with the main aim to detect as many real
4.5\,$\mu$m sources as possible. At the faint end (below 20\,$\mu$Jy)
the photometric uncertainties may become large; even for nominal
5-$\sigma$ detected sources the real uncertainties may exceed 20\%, in
particular for the other IRAC bands
(see Fig.~\ref{fig_comparison_seip}). 
Any photometric errors will smear out the color contrasts needed and thus
will lower the detection of galaxy overdensities.  
Nevertheless the search for galaxy overdensities turns out to be
successful (Sec.~\ref{sec:color_cuts}).
Thereby the galaxy overdensities are based on typically a dozen
galaxies, the brightest cluster members.
While this search only uses the 3.6 and 4.5\,$\mu$m data of the
central 3C fields (and the corresponding $i$ band data), for future purposes 
we have created the catalog also for the other
filters and the off-fields.
The catalog can be obtained from the authors on request.

For the rest of the paper, we removed some extreme sources from the catalog:
1) bright sources ($\sim$1\,\%) 
having F(4.5$\mu$m) $>$ 1000 $\mu$Jy (16.4 AB mag), 
about 3.5 mag brighter than an $L^{\star}$ galaxy at $z>1$
\citep{Wylezalek14}. 
2) We also removed extended sources ($\sim$15\,\%) where the IRAC Kron
semimajor axis is larger than 3$\farcs$8. At $z>1$ a galaxy of this size
has a diameter of about 60\,kpc and we do not expect it to be more
extended.
The size cut also removes IRAC double sources which have not properly
been deblended.
3) Finally, we removed all IRAC sources with a double or multiple  PSO
counterpart, to avoid color mismatch. We do not expect that this source removal creates galaxy overdensities.

\subsection{Foreground galactic
  extinction}
\label{sec:fge}

\begin{table*}
\renewcommand{\thetable}{\arabic{table}}
\centering
\caption{Galactic coordinates of the 3C sample and foreground galactic
  extinction at the $i$ band and  IRAC1 and IRAC2 
  (in magnitudes) taken from NED, based on \citet{Schlafly11}.
} 
\label{table_fge}
\begin{tabular}{lrrrrr|lrrrrr}
\hline
\hline
Name         &  longitude &   latitude & $i$      & IRAC1  & IRAC2  &  Name         &  longitude &   latitude &  $i$     & IRAC1  & IRAC2   \\ 
\hline                                                                         
3C\,002      &    99.2804 &   -60.8588 & 0.078    & 0.008  & 0.007  &  3C\,250      &   212.3735 &    66.9100 &  0.028   & 0.003  & 0.002   \\
3C\,009      &   112.0466 &   -46.5331 & 0.089    & 0.009  & 0.008  &  3C\,252      &   184.8085 &    67.1171 &  0.035   & 0.004  & 0.003   \\
3C\,013      &   119.3154 &   -23.3473 & 0.086    & 0.009  & 0.008  &  3C\,255      &   263.0142 &    52.5427 &  0.092   & 0.010  & 0.008   \\
3C\,014      &   117.8728 &   -44.0881 & 0.112    & 0.012  & 0.010  &  3C\,256      &   218.0286 &    69.2001 &  0.029   & 0.003  & 0.003   \\
3C\,036      &   127.7840 &   -17.0153 & 0.110    & 0.012  & 0.010  &  3C\,257      &   254.8015 &    59.8493 &  0.063   & 0.007  & 0.006   \\
3C\,043      &   134.2174 &   -38.3866 & 0.152    & 0.016  & 0.013  &  3C\,266      &   147.6420 &    64.0886 &  0.031   & 0.003  & 0.003   \\
3C\,065      &   141.4979 &   -19.5069 & 0.084    & 0.009  & 0.007  &  3C\,267      &   254.8051 &    69.6831 &  0.058   & 0.006  & 0.005   \\
3C\,068.1    &   145.6014 &   -23.9886 & 0.116    & 0.012  & 0.010  &  3C\,268.4    &   147.4838 &    71.4045 &  0.024   & 0.002  & 0.002   \\
3C\,068.2    &   147.3258 &   -26.3775 & 0.230    & 0.024  & 0.020  &  3C\,270.1    &   166.3069 &    80.6393 &  0.021   & 0.002  & 0.002   \\
3C\,119      &   160.9652 &    -4.3423 & 0.952    & 0.100  & 0.083  &  3C\,280.1    &   115.2599 &    76.8402 &  0.028   & 0.003  & 0.002   \\
3C\,124      &   195.5094 &   -27.7441 & 0.206    & 0.022  & 0.018  &  3C\,287      &    22.4661 &    80.9884 &  0.019   & 0.002  & 0.002   \\
3C\,173      &   178.9788 &    18.2911 & 0.177    & 0.019  & 0.015  &  3C\,294      &    61.1329 &    72.3714 &  0.024   & 0.003  & 0.002   \\
3C\,181      &   203.7542 &    14.6299 & 0.107    & 0.011  & 0.009  &  3C\,297      &   339.9002 &    52.5772 &  0.091   & 0.010  & 0.008   \\
3C\,186      &   181.7870 &    26.1410 & 0.085    & 0.009  & 0.007  &  3C\,298      &   352.1592 &    60.6666 &  0.050   & 0.005  & 0.004   \\
3C\,190      &   207.6235 &    21.8410 & 0.052    & 0.006  & 0.005  &  3C\,300.1    &   346.1491 &    53.1231 &  0.084   & 0.009  & 0.007   \\
3C\,191      &   211.8721 &    20.9011 & 0.038    & 0.004  & 0.003  &  3C\,305.1    &   114.9214 &    38.3324 &  0.050   & 0.005  & 0.004   \\
3C\,194      &   177.8578 &    31.9170 & 0.083    & 0.009  & 0.007  &  3C\,318      &    29.6412 &    55.4156 &  0.107   & 0.011  & 0.009   \\
3C\,204      &   150.3443 &    35.5122 & 0.134    & 0.014  & 0.012  &  3C\,322      &    88.5228 &    49.1338 &  0.023   & 0.002  & 0.002   \\
3C\,205      &   159.2604 &    36.8964 & 0.133    & 0.014  & 0.012  &  3C\,324      &    34.9362 &    49.1595 &  0.079   & 0.008  & 0.007   \\
3C\,208      &   213.6616 &    33.1581 & 0.057    & 0.006  & 0.005  &  3C\,325      &    96.2666 &    44.0857 &  0.023   & 0.002  & 0.002   \\
3C\,208.1    &   213.5991 &    33.5803 & 0.051    & 0.005  & 0.004  &  3C\,326.1    &    33.6858 &    47.3295 &  0.086   & 0.009  & 0.007   \\
3C\,210      &   197.7630 &    38.7775 & 0.052    & 0.005  & 0.005  &  3C\,356      &    77.9218 &    34.2051 &  0.054   & 0.006  & 0.005   \\
3C\,212      &   213.9971 &    34.5042 & 0.069    & 0.007  & 0.006  &  3C\,368      &    37.7076 &    15.2246 &  0.250   & 0.026  & 0.022   \\
3C\,220.2    &   188.1127 &    46.7802 & 0.026    & 0.003  & 0.002  &  3C\,418      &    88.8081 &     6.0405 &  1.784   & 0.187  & 0.155   \\
3C\,222      &   230.1417 &    38.3249 & 0.077    & 0.008  & 0.007  &  3C\,432      &    67.9354 &   -22.8248 &  0.152   & 0.016  & 0.013   \\
3C\,225      &   219.8665 &    44.0246 & 0.092    & 0.010  & 0.008  &  3C\,437      &    70.8753 &   -28.3921 &  0.151   & 0.016  & 0.013   \\
3C\,230      &   237.5733 &    39.0949 & 0.128    & 0.013  & 0.011  &  3C\,454.0    &    87.3528 &   -35.6483 &  0.098   & 0.010  & 0.009   \\
3C\,238      &   234.0453 &    46.6644 & 0.037    & 0.004  & 0.003  &  3C\,454.1    &   113.5670 &    10.8496 &  0.689   & 0.072  & 0.060   \\
3C\,239      &   170.4717 &    53.1961 & 0.015    & 0.002  & 0.001  &  3C\,469.1    &   120.3877 &    17.3280 &  0.383   & 0.040  & 0.033   \\
3C\,241      &   213.2053 &    55.7394 & 0.045    & 0.005  & 0.004  &  3C\,470      &   113.0052 &   -17.7716 &  0.181   & 0.019  & 0.016   \\
3C\,245      &   233.1245 &    56.3004 & 0.042    & 0.004  & 0.004  &  4C\,13.66    &    40.0101 &    17.1849 &  0.295   & 0.031  & 0.026   \\
3C\,249      &   255.6952 &    51.2821 & 0.071    & 0.007  & 0.006  &  4C\,16.49    &    39.3069 &    24.0067 &  0.125   & 0.013  & 0.011   \\
\hline
\end{tabular}
\end{table*}

The 3C sample is selected at low radio frequency 178 MHz
\citep{Spinrad85}. At this freuency any 
foreground galactic extinction is negligible. Therefore the 3C sources
are found also at low galactic latitude, where foreground galactic
extinction may become relevant at optical wavelengths. 
Table~\ref{table_fge} lists the galactic coordinates as well as the
foreground galactic 
extinction in those three bands used for the galaxy overdensity search
($i$,IRAC1 and IRAC2).

Among the 64 sources six sources lie close ($\lesssim 15^{\circ}$) to
the galactic plane.  
These are
3C\,119, 3C\,368, 3C\,418, 3C\,454.1,3C\,469.1, 4C\,13.66.
While their foreground galactic
extinction is negligible in the IRAC bands, 
in the $i$ band it is substantial with values 
in the range $0.25 < A_{i} < 1.8$. 
For most of the remaining 3C sources the  $i$ band foreground galactic
extinction is small with respect to the $i - [4.5]$ color uncertainties.

For the rest of the paper, in particular for the galaxy overdensity
search, {\it all sources are corrected for foreground galactic 
extinction} using the values of Table~\ref{table_fge}. 
If a source is actually a star in the Milky Way, then its extinction
may be lower so that the 
correction makes it too blue but a star shall be removed anyway by 
the color cuts applied.

\section{SED examples} \label{sec:sed_examples}

\begin{figure*}
  \includegraphics[width=16.5cm, clip=true]{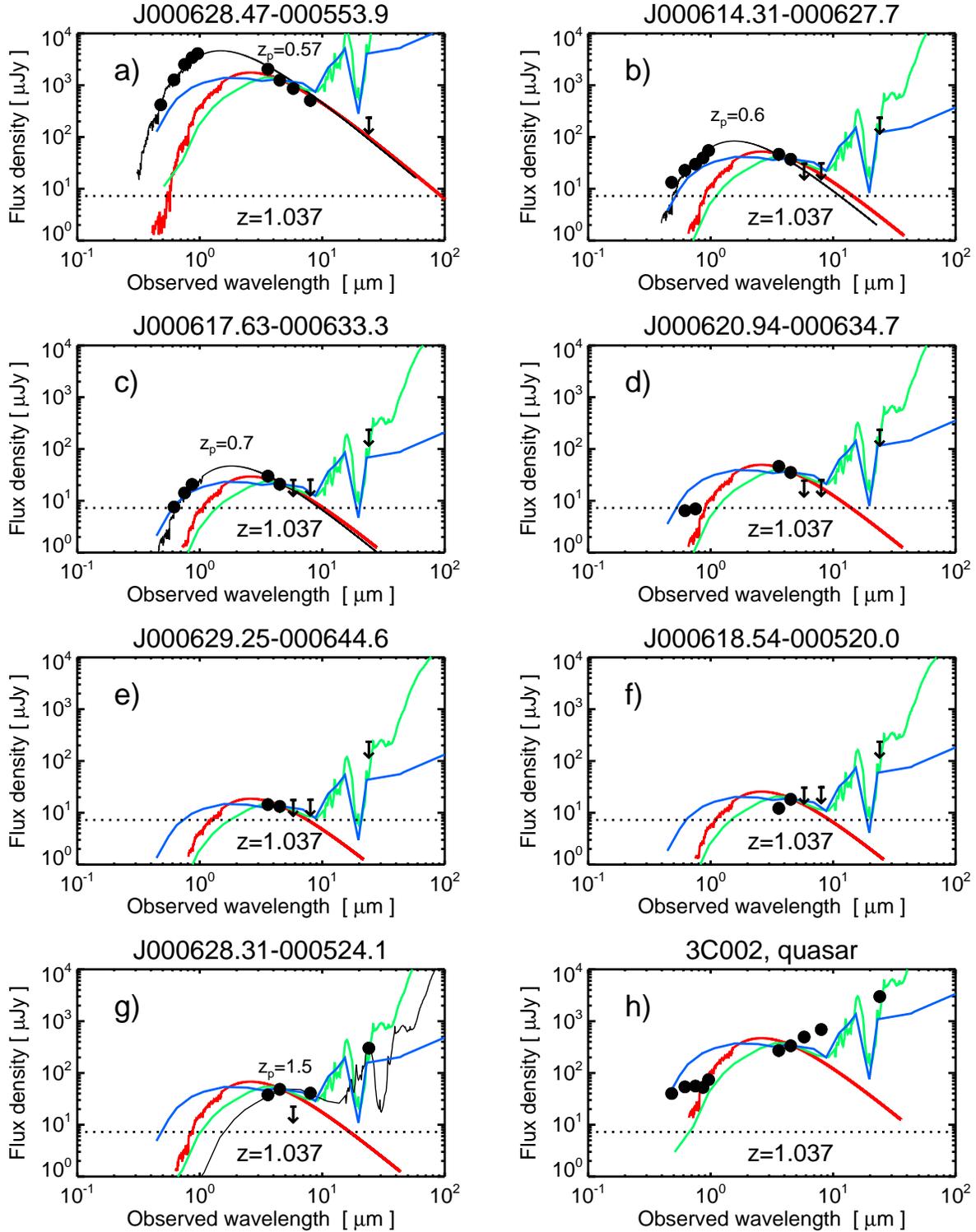}
  \caption{Eight example SEDs of galaxies in the field of the quasar
    3C\,002.
    The SEDs
    shown are typical for the entire sample. 
    Black dots denote detections, black arrows denote 3-$\sigma$ upper
    limits, and error bars are smaller than the symbol
    size.
    The dotted horizontal line marks  the Pan-STARRS $i$-band detection limit
    (7.2\,$\mu$Jy).
    The first four examples (a$-$d) are detected with both {\it Spitzer} and
    Pan-STARRS, the next three examples (e$-$g) only with {\it
      Spitzer}. The
    last example (h) shows the quasar itself with a power-law SED, so that none of the galaxy template SEDs fits.
    The three colored solid lines illustrate galaxy templates shifted to
    the redshift $z=1.037$ of 3C\,002:
    red = elliptical (NGC\,221), 
    green = ultraluminous infrared galaxy (ULIRG, Arp\,220), 
    blue = star forming spiral galaxy (NGC\,6946).
    For panels a), b), c) the black line shows the elliptical template at the
    best fitted photometric redshift $z_{\rm p}$, for panel g) a
    background ULIRG at $z_{\rm p} = 1.5$ fits bests.
  }
  \label{fig_seds_1}
\end{figure*}

Figure~\ref{fig_seds_1} depicts the spectral energy distributions of
eight typical galaxies in the field of the quasar 3C\,002 along with
three galaxy SED templates. The three templates were chosen to embrace
the range of candidate 
cluster member galaxies: NGC\,6946 is a relatively blue star-forming
spiral galaxy with 
moderate FIR luminosity ($L_{\rm FIR} < 10^{11} L_{\odot})$, Arp\,220 is
an archtypal ultraluminous infrared galaxy (ULIRG), 
and NGC\,221 is an old elliptical with marginal post-starforming
signatures. The template SEDs were taken from the NASA Extragalactic
Database (NED)\footnote{http://ned.ipac.caltech.edu/}.

The important features of each Figure~\ref{fig_seds_1} example are respectively: 

\begin{figure*}
  \includegraphics[width=8cm, clip=true]{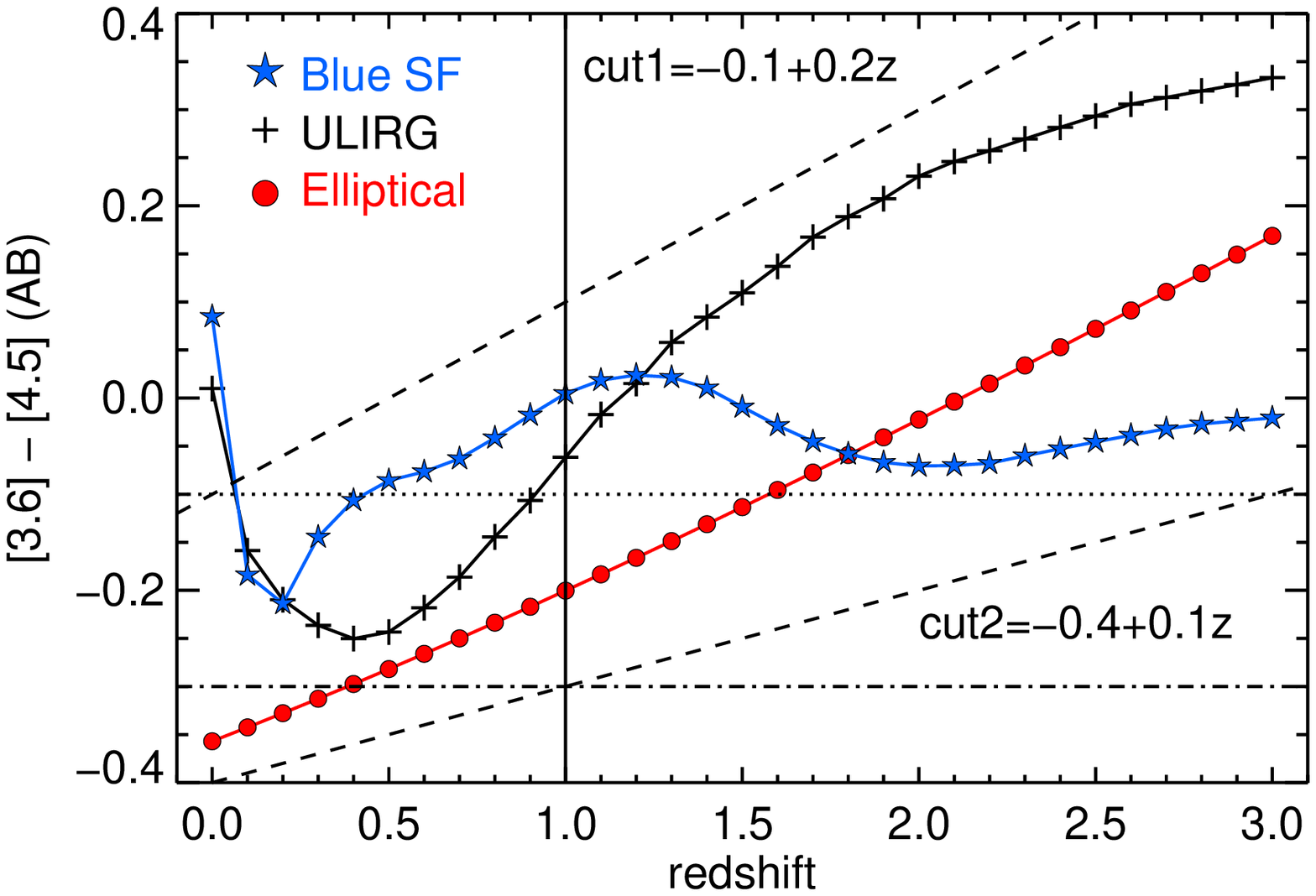}
  \includegraphics[width=8cm, clip=true]{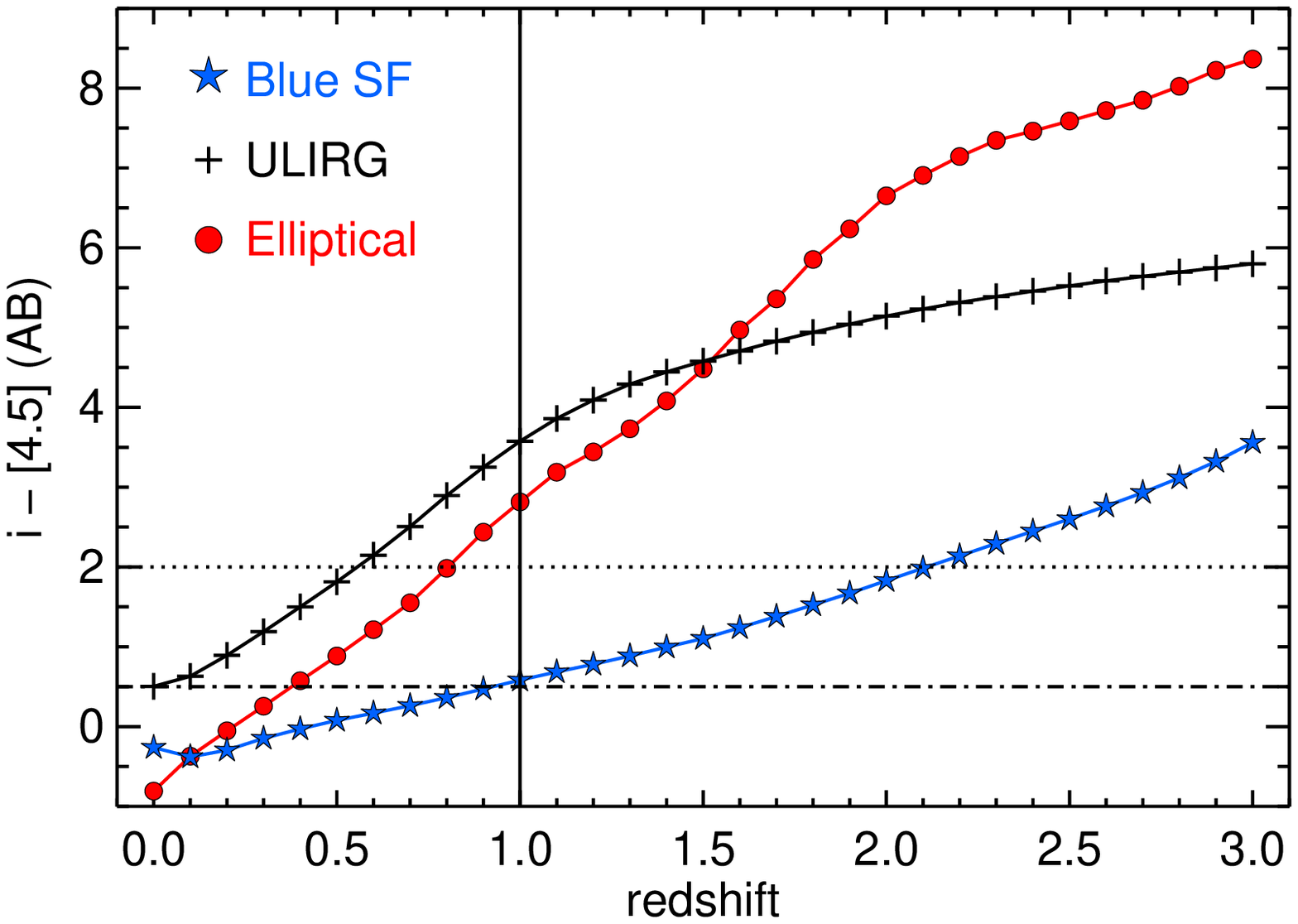}
  \caption{Color $[3.6] - [4.5]$ (left)
    and $i - [4.5]$ (right)
    versus redshift for three templates:  elliptical (NGC\,221), 
    ultraluminous infrared galaxy (ULIRG, Arp\,220), 
    blue star forming spiral galaxy (Blue SF, NGC\,6946).
    The black dotted, dash-dotted and solid lines mark
    color and redshift cuts 
    discussed in Section~\ref{sec:color_cuts}.
    The long-dashed slopes in panel (a) embrace a redshift dependent color
    range between cut1 and cut2.
  }
  \label{fig_template_color_vs_redshift}
\end{figure*}

\begin{itemize}
\item[a)] Bright, about a factor of 5 above the IRAC completeness
  limits (36\,$\mu$Jy). 
  The Pan-STARRS -- {\it Spitzer} color is $i - [4.5] \ll 0$ AB mag.
  This source is a foreground elliptical at photometric redshift
  $z_{\rm p} =0.57$.  

\item[b)] Pan-STARRS $i$ similarly bright as 
  {\it Spitzer}, $i - [4.5] \sim 0$. This source is a
  foreground elliptical at $z_{\rm p} =0.6$. 

\item[c)] Pan-STARRS slightly fainter than {\it Spitzer}, Pan-STARRS SED
  very red, only
  detected in the $r$, $i$ and $z$ bands, $i - [4.5] \sim 0$. 
  A foreground elliptical at $z_{\rm p} =0.7$ fits best.

\item[d)] Pan-STARRS much fainter than {\it Spitzer}, only
  detected in the $r$ and $i$ bands, $i - [4.5] \sim 2$.  
  The SED may belong to a star forming galaxy type between the NGC\,6946
  and Arp\,220 templates at $z=1.037$ of 3C\,002.

\item[e)] Not detected in Pan-STARRS; detected in only two IRAC
  bands. 
  $[3.6] - [4.5] \sim -0.09$. Each
  template fits well at the redshift of 3C\,002, but 
  the non-detection at $i$ may reject a blue SF galaxy at $z=1.037$.

\item[f)] Not detected in Pan-STARRS; detected in only two IRAC
  bands.
  $[3.6] - [4.5] \sim 0.44$. Potentially a 
  background ($z > 3$) galaxy with intrinsically red color.

\item[g)] Not detected in Pan-STARRS, 
  detected in three IRAC bands, 
  $[3.6] - [4.5] \sim 0.3$,  and in in MIPS. 
  Best fitted by 
  a background ULIRG at  $z_{\rm p} =1.5$. This is consistent with the
  detection on {\it Herschel} PACS and SPIRE maps (Westhues et al.\ in
  prep). The Pan-STARRS non-detection
  rejects a blue star forming galaxy at $z \sim 1.0$. 

\item[h)] The quasar 3C\,002, detected in all
  bands with a power-law rising infrared SED typical for type-1
  AGN. $[3.6] - [4.5] \sim 0.3$, $i - [4.5] \sim 2$.
\end{itemize}
The SED examples demonstrate that the combination of Pan-STARRS and
{\it Spitzer} constrains the redshifts of
galaxies, both for Pan-STARRS detections and non-detections.

\section{Color cuts} \label{sec:color_cuts}

To make the search for candidate cluster member galaxies in the high-$z$ 
3C environments more sensitive, color cuts have to be
defined. 
Ideally such cuts should remove as many foreground stars and galaxies as
posssible but none or as few as possible of the candidate cluster
member galaxies.  
In practice, for data with uncertainties in the color, a compromise has
to be chosen. 
Figure~\ref{fig_template_color_vs_redshift} shows the color 
as a function of redshift for the three templates introduced in the
previous section. 
We here consider two colors, a) the IRAC1/2 color  $[3.6] - [4.5]$
and b) the  Pan-STARRS$-$ IRAC color $i - [4.5]$. 
The pure IRAC color  $[3.6] - [4.5]$ has successfully been applied to
other data sets, e.g., \citet{Wylezalek13}. 
The combination of IRAC and optical colors has also been applied in
the past on other data sets to remove foreground sources and to
identify candidate high-$z$ 
galaxies and clusters, e.g., by \citet{Eisenhardt04, Eisenhardt08} and
\citet{Falder11} with optical data from 
the NOAO and INT, respectively.

IRAC1/2:
  For $z \lesssim 1$ the three templates (with added uncertainties)
  populate the color range $-0.4 < [3.6] - [4.5] < 0.1$.
  At  $z \gtrsim 1$, 
  $[3.6] - [4.5] > -0.1$
  selects blue star
  forming (SF) galaxies and ULIRGs and presumably any star
  forming galaxy.
  It appears most efficient to exclude foreground ULIRGs at $0.1 < z <
  1$ and SF at $0.1 < z < 0.5$.
  For comparison, 
  based on synthetic galaxy models by \citet{Bruzual03},
  Wylezalek et al. (2013, see their Fig.~3) 
  used this same color cut to select
  essentially all types of galaxies at  $z \gtrsim 1.3$.
  However, this color cut excludes also 
  ellipticals at $z < 1.5$, i.e., the redshift of most of the 3C sources.
  Therefore, we will here use the color cut 
  $[3.6] - [4.5] > -0.3$, which excludes many but not all $z<1$
  galaxies. It also excludes stars all having $[3.6] - [4.5] \sim
  -0.5$ ($\sim$\,0 Vega mag).
  
PSO--IRAC:
  The color $i - [4.5]$ shows a steep redshift dependence for elliptical
  and red starforming galaxies (ULIRGs), rising between $z=1$ and
  $z=2.5$ from $i - [4.5] \sim  3$  to  $i - [4.5] \sim 6$.
  For blue star forming galaxies the color rises more slowly from 
  $i - [4.5] \sim  0.5$  to  $i - [4.5] \sim 2.5$. 
  To exclude $z<1$ foreground galaxies a color cut $i - [4.5] >2$
  appears efficient, but it will also exclude blue SF galaxies at $1<z<2$.
  Therefore, we will here use the softer color cut 
  $i - [4.5] > 0.5$, which again should be reliable at the expense of
  letting though some $z<1$ galaxies. 
  This cut also excludes most stars
  ($i - [4.5] \sim  0.5$ AB mag corresponds to $R_{J} - [4.5] \sim$ 3.5 Vega mag).

The main difference between the IRAC1/2 and PSO--IRAC color cuts is that the IRAC1/2
color requires high photometric accuracy, while
the range of the PSO--IRAC colors is so large that even a color
uncertainty of 0.3 mag plays a minor role.
Also, at  $1<z<1.4$ the galaxy types close to the color cuts
differ, being passive ellipticals for the IRAC cut but blue
star-forming galaxies for the PSO-IRAC cut.

\begin{figure}
  \includegraphics[width=8cm, clip=true]{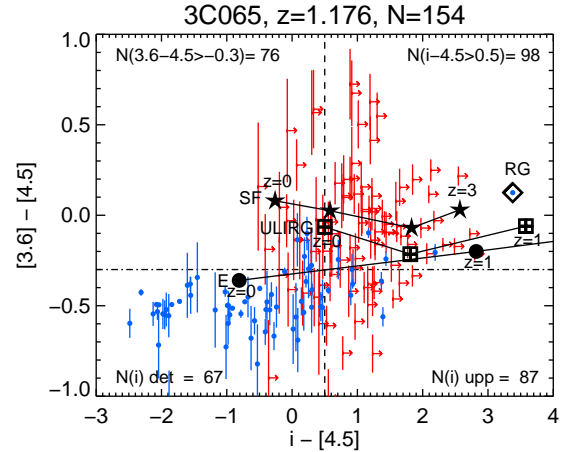}
  \caption{Color-color diagram $[3.6] - [4.5]$ versus $i - [4.5]$
    of the 3C\,065 field typical of the entire sample.
    Only sources with  $F(4.5) > 4 \mu$Jy and located within 120$\arcsec$
    angular separation from the 3C source are plotted.
    Blue dots and  red arrows with vertical
    error bars distinguish between  $i$ band detected  and
    undetected sources, respectively.
    The lower limits in $i - [4.5]$ assume $i = 21.7$ AB mag
    (Tab.~\ref{tab_detection_limits}). 
    The redshift dependent positions of the three templates (SF, ULIRG
    and Elliptical) are shown, as well as the position of the 3C AGN
    itself (diamond, RG).
    Note that stars lie around $[3.6] - [4.5] = -0.5$ (Vega color
    $\sim$ 0).
    The vertical dashed and horizontal dashed-dotted lines mark the
    adpted color
    cuts.
    The number of sources is given for the $i$ band
    detections, non-detections, and for both cuts.
  }
  \label{fig_ccd}
\end{figure}

To further justify the choice of the color cuts, 
we consider color-color
diagrams of our data, taking into account 
also the $i$ band upper limits. 
Figure~\ref{fig_ccd} shows an example representative for the entire sample. 
The striking features of
the color-color diagrams are:

\begin{enumerate}
\item 
  The removal of foreground stars and galaxies is essential to reveal candidate 
  cluster member galaxies.
  The foreground sources  are bluer than both 
  the IRAC1/2 and the PSO--IRAC color cuts.

\item
  Each of the color
  cuts removes about 70\%  of the $i$ band detections.
  A stricter IRAC color cut $[3.6] - [4.5] > -0.1$ or 
  a stricter PSO--IRAC color cut $i - [4.5] > 2$ would remove nearly
  all $i$ band detections. 
  However, the $i$ band non-detections remain potential 
  cluster member galaxies because of the lower limit nature of their
  color.

\item
  Both color cuts $[3.6] - [4.5] > -0.3$ and $i - [4.5] > 0.5$ 
  select also high-$z$ AGN, as indicated by the 3C
  source itself.

\item
  For about half of our sources, 
  the uncertainty of the IRAC1/2 color exceeds
  0.1 mag. This is large relative to the IRAC color range.
  The relatively large uncertainty of the IRAC1/2 color makes it
  likely that a stricter IRAC1/2 color cut will
  reject also candidate cluster member galaxies.
  On the other hand, for all sources 
  the uncertainty of the PSO--IRAC color is small enough ($<$0.3\,mag) to be
  negligible at the PSO--IRAC color range of several mag. 

\item
  On average, only about 40\% of the sources 
  have an $i$ band detection, hence a measured PSO--IRAC color; the
  remaining 60\% of the sources with $i$-band upper limits 
  have only a lower limit for PSO--IRAC color.

\item
  On average, the IRAC1/2 and PSO--IRAC color cuts reject 45\%
  and 30\%  of the sources, respectively.
\end{enumerate}

We have tested several color cuts on the present 3C data, i.e., 
including redshift-dependent and redshift-independent ones, using 
colors based on other Pan-STARRS
filters and 
a criterion accepting only the $i$-band upper limits as candidate cluster
member galaxies. 
These tests led us to the redshift-independent color cuts either 
$[3.6] - [4.5] > -0.3$  or
$i - [4.5] > 0.5$ plus all $i$-band
non-detections.
For the present purpose, sources satisfying either of these cuts are  
considered high-$z$ candidate galaxies.

\section{Galaxy overdensities} \label{sec:galaxy_overdensities}

\subsection{Radial surface density profiles}
\label{sec:density_profiles}

\begin{figure}
  \includegraphics[width=4cm, clip=true]{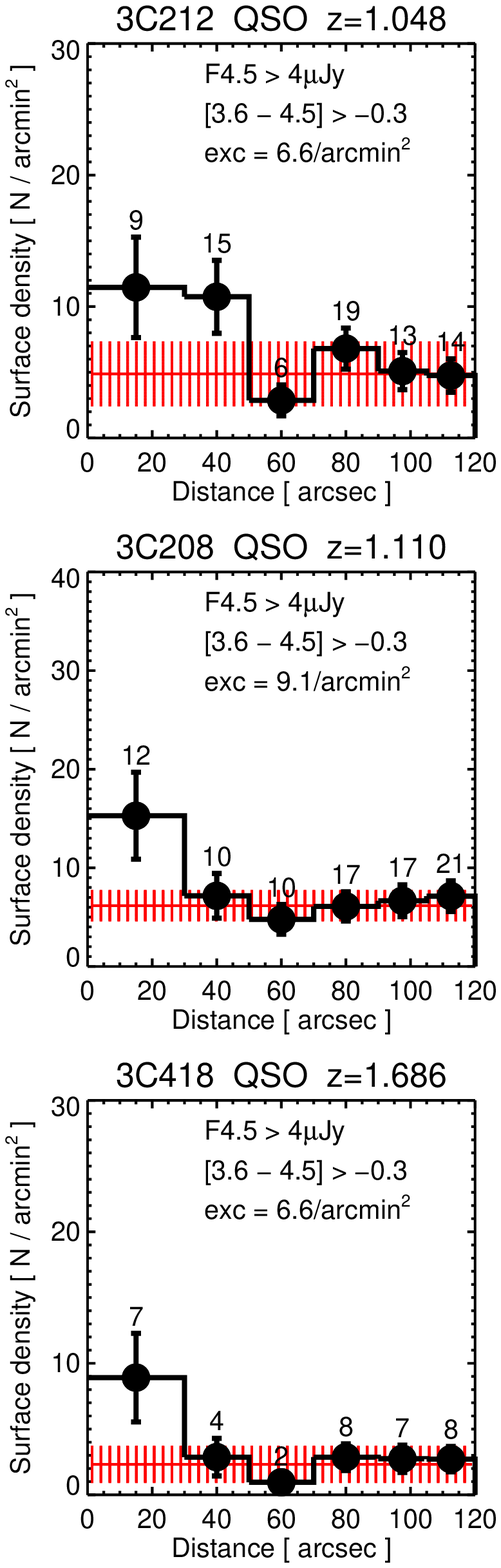}
  \includegraphics[width=4cm, clip=true]{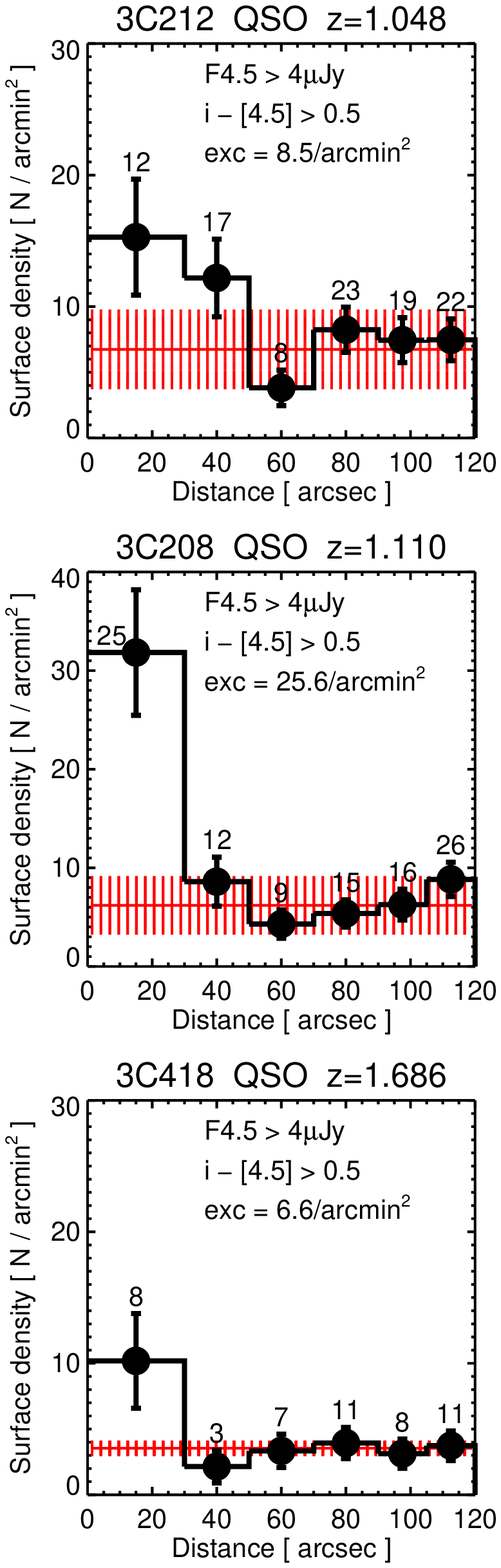}
  \caption{Radial surface density of
    candidate cluster member galaxies for three 3C fields with a clear central overdensity.
    The candidate cluster member galaxies were selected with 
    (a) the IRAC color
    cut $[3.6] - [4.5] > -0.3$ AB mag 
    (left column) or (b) the Pan-STARRS$-$IRAC criterion (right column). 
    The bin radii are 30, 50, 70, 90, 105, 120$\arcsec$.
    30$\arcsec$ corresponds to about 250\,kpc projected separation 
    for all redshifts of this work. 
    Numbers show
    the number of candidates in each radial bin. 
    Error bars show Poisson
    uncertainty in each bin. 
    The red shaded areas mark the $\pm$3-$\sigma$ 
    range around the mean
    surrounding density (red solid line),
    determined from the 4 outermost annuli at $50\arcsec - 120\arcsec$. 
    The excess density as labeled 
    gives the density  inside 30$\arcsec$
    above the mean surrounding density.
    The 3C source
    is not counted in the radial density plots.
    The examples are typical for those 3Cs with a clear central overdensity.
  }
  \label{fig_single_radial_densities}
\end{figure}

For each 3C source\footnote{Three 3C sources are
excluded from the analysis here:
3C\,239 has a poor S/N on the IRAC 4.5\,$\mu$m map, 
and both 3C\,294 and 3C\,470 are contaminated by a bright
nearby star preventing the detection of candidate cluster member 
galaxies out to 15$\arcsec$ radius. 
Furthermore, 3C\,294 lies less than 2$\arcmin$ from the IRAC map border,
preventing the proper determination of the surrounding density.
} 
we determined the radial surface density of galaxies satisfying 
color cuts of Section~\ref{sec:color_cuts}. 
These galaxies are considered candidate
cluster members. 
We applied two methods to identify the candidate cluster member 
galaxies:
\begin{enumerate}
\item The IRAC color selection 
$[3.6] - [4.5] > -0.3$ AB mag. 

\item The PSO$-$IRAC criterion introduced here:
   $i - [4.5] > 0.5$ AB mag plus all $i$-band
   non-detections.

\end{enumerate}
For both methods we applied a flux cut
requiring $F4.5 >$~4\,$\mu$Jy (22.4 AB mag).
For comparison, from the clustering analysis of the CARLA sample, 
\citet{Wylezalek14} found that 
the 4.5\,$\mu$m magnitude of an $L^{\star}$ galaxy in the redshift
range
$1.3 < z < 3.1$ is about 20 AB mag.
Figure~\ref{fig_single_radial_densities} depicts radial surface
density profiles 
of
candidate cluster member galaxies for three 3C fields with a clear
central overdensity.

\begin{figure}
  \includegraphics[width=4cm, clip=true]{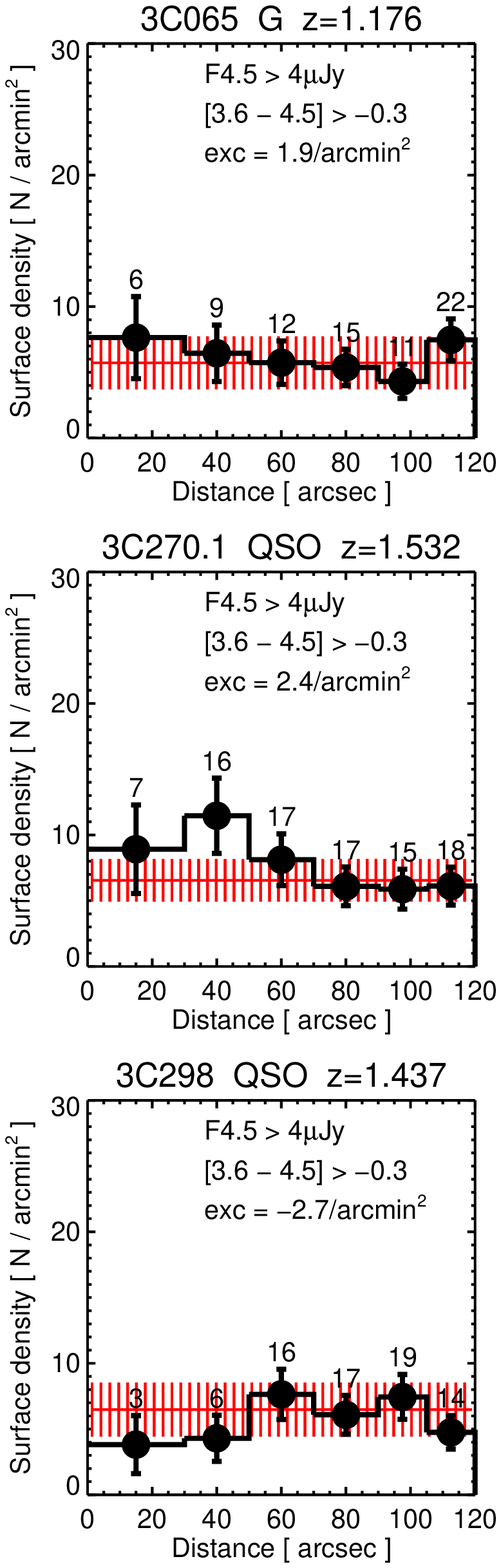}
  \includegraphics[width=4cm, clip=true]{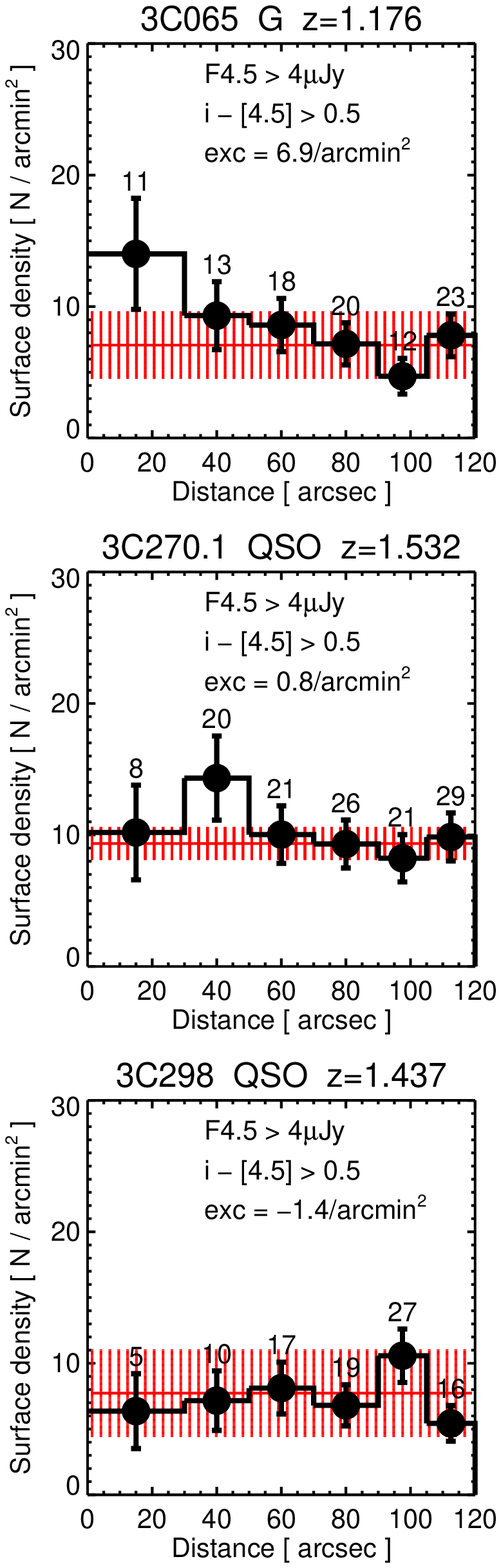}
  \caption{Same as Figure \ref{fig_single_radial_densities} but for 3C fields
    with less pronounced or no central overdensity.
  }
  \label{fig_single_radial_densities_without_overdensity}
\end{figure}

For either selection criterion the radial surface density  of 
candidate cluster member galaxies varies from field to field.  
For each 3C field we determined the central 
radial surface density ($D_{cent}$) within 30$\arcsec$  
and the mean surrounding density ($D_{surr}$)
between 50$\arcsec$ and 120$\arcsec$ and the uncertainty of the 
mean ($U_{surr}$).\footnote{We did not use the
  IRAC off fields to determine tha surrounding because these fields
  were observed in two IRAC bands only hence have 
  brighter completeness and detection limits than the central field
  observed in four IRAC bands.}
Using $Exc = D_{cent} - D_{surr}$,
we define overdensity as 
excess $Exc$ of the central 
radial density above the mean surrounding density.
We consider the overdensity as significant, 
if $Exc$ lies 
above the mean surrounding density by three times the uncertainty, i.e. if 
$Exc > D_{cent}  - (D_{surr} + 3 \cdot U_{surr})$.
The numbers of candidates in each radial bin 
are listed for both color selection methods 
in Table~\ref{table_radial_densities}.
Table~\ref{table_overdensities} list the central and average 
surrounding surface densities.

About 50\% of the 3C sources show an overdensity of candidate cluster member
galaxies within the  
central 30$\arcsec$ (and 20\% of them even within 30--50$\arcsec$), 
and the remaining $\sim$50\% do not. 
While the examples of 
Figure~\ref{fig_single_radial_densities} exhibit a clear central galaxy
overdensity, Figure~\ref{fig_single_radial_densities_without_overdensity}
shows examples with less pronounced or no
overdensity.
Inspection of the IRAC maps confirms that in those cases with no
overdensity there are, in fact, very few sources near the radio
source. 
The surrounding density varies considerably between the 3C fields, 
and within a field (outside of the central area) the distribution
of candidate cluster member galaxies is not homogeneous 
("cosmic variance"). 
Despite the  variations, both color selection methods yield
remarkably consistent overdensity results:  
if for a given 3C field one method finds a central overdensity, 
then mostly the other method does as well (Tab~\ref{table_overdensities}).

\subsection{Statistics of the overdensities}
\label{sec:statistics}

\begin{figure}
  \includegraphics[width=8cm, clip=true]{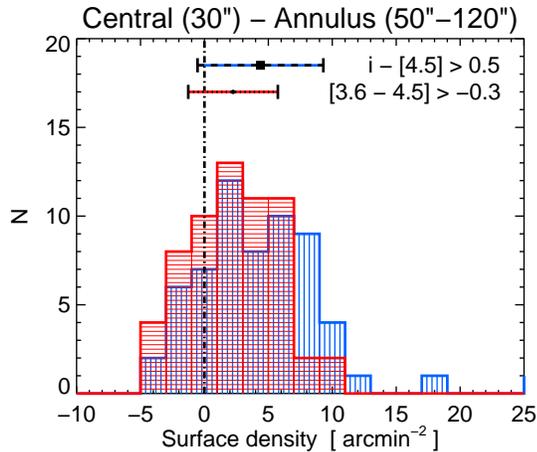}
  \caption{Distribution of the central
    overdensities, i.e., central surface density minus surrounding surface
    density, for both color selections  IRAC1/2 (red, dotted)
    and  PSO$-$IRAC (blue, dashed).
    For both distributions the average with standard
    deviation are plotted above the histograms.
    The uncertainty of the mean is smaller than the size of the square. 
    The vertical black dash-dotted line marks zero excess surface density.
  }
  \label{fig_density_histogram}
\end{figure}

Figure~\ref{fig_density_histogram} shows the distribution 
of the central 30$\arcsec$ overdensities 
(i.e., positive and negative excess) over the 50$\arcsec$--120$\arcsec$ surrounding 
as measured by both the IRAC1/2 and the PSO--IRAC color
cuts for the 61 ~ 3C fields used.
Both color cuts yield, on average, an overdensity with values for average
and standard deviation of 
2.2 $\pm$ 3.5 (IRAC1/2) and 4.4 $\pm$ 4.9 (PSO--IRAC). 
The uncertainties of the mean ($\sigma$/sqrt[61]) are 0.45 and 0.63,
respectively, and show that the
means are $>$\,5$\sigma$ above zero. 
However, the
distributions are broad and include zero, indicating that 
some 3C fields have an overdensity and others do not, consistent
with the findings shown in Figures~\ref{fig_single_radial_densities} 
and \ref{fig_single_radial_densities_without_overdensity}.

\begin{figure}
  \includegraphics[width=8cm, clip=true]{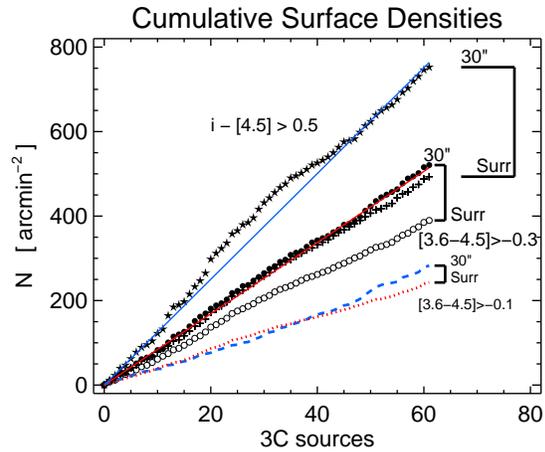}
  \caption{Cumulative surface densities
    of the central bin (30$\arcsec$
    radius) and the mean surrounding 
    for the IRAC color selection (filled and open
    circles, respectively)
    and the  PSO$-$IRAC color selection  (stars and crosses, respectively).
    The X-axis refers to the index of the 3C sources sorted by
    redshift, i.e., $z\approx1.0$ at X=1, $z\approx1.4$ at X=35, $z\approx1.6$ at X=45, and $z\approx2.5$ at X=61.
    The blue and red solid lines show a linear fit through the PSO$-$IRAC
    central and surrounding distributions, respectively.
    For comparison, 
    the blue dashed and red dotted lines with small labels show the
    results for a strong IRAC1/2 cut  $[3.6] - [4.5] > -0.1$
  }
  \label{fig_cumulative_densities}
\end{figure}

To further check the statistical significance of the mean central
overdensities, we consider the cumulative surface density
distributions as shown in Figure~\ref{fig_cumulative_densities}.
For both color cuts, the cumulative distribution of the central bin
diverges continuously from the cumulative average surrounding. 
For both color cuts, the  two-sided KS test gives an extremely low
probability that the measured central and surrounding 
distributions are drawn from the same intrinsic populations. The
probabilities of finding such large deviations are
10$^{-4.7}$ (IRAC) and 10$^{-6.8}$ (PSO--IRAC).
This provides clear statistical evidence for the average central
overdensities.
As already indicated by 
Figure~\ref{fig_density_histogram}, 
the PSO-IRAC selection method reaches a 
higher significance 
for overdensities than the
IRAC1/2 selection.

The surrounding counts shown in Figure~\ref{fig_cumulative_densities}  
ought to be linearly rising with increasing X, which is the index of the
3C sources sorted by redshift.  The surrounding
galaxies do not care whether there is a 3C source nearby or what its
redshift is.\footnote{An exception may occur if strong cluster lensing
plays a role, but this topic is beyond the scope of the current study.} 
In fact, the surrounding counts of the PSO--IRAC selection follow a
linear relation, while the cumulative central counts of the PSO--IRAC selection
show a clear deviation from the linear fit, i.e., a trend of a decline
of overdensities with increasing redshift.
On the other hand, both surrounding and central counts of the IRAC
selection show a good linear relation with X, and any
deviation is marginal.

\begin{figure}
  \includegraphics[width=8cm, clip=true]{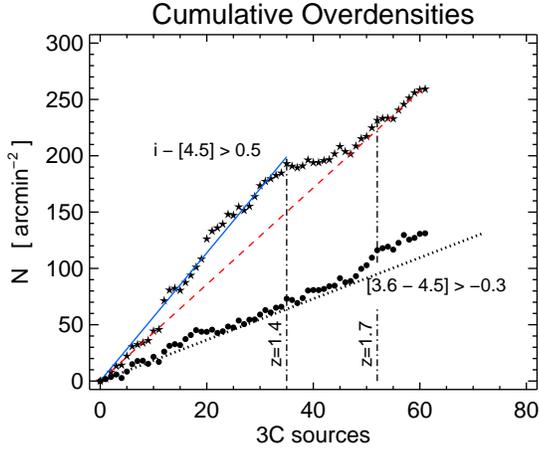}
  \caption{Cumulative overdensities
    for the IRAC color selection $[3.6] - [4.5] > -0.3$ (circles)
    and the PSO$-$IRAC color selection (stars).
    The overdensities are the central surface densities  
    after subtraction of the mean surrounding (see Fig.~\ref{fig_cumulative_densities}).
    The X-axis refers to the index of the 3C sources sorted by
    redshift.
    The vertical dash-dotted lines mark the X-position of 3C sources at
    $z=1.4$ and $z=1.7$.
    The dotted line shows the fit of the IRAC slope, whereby  the
    range near $z=1.7$ was excluded. 
    The solid blue and dashed red lines embrace the range of the PSO$-$IRAC slopes.
  }
  \label{fig_cumulative_densities_2}
\end{figure}

To demonstrate the redshift trends more clearly, 
Figure~\ref{fig_cumulative_densities_2} shows the cumulative
overdensities, i.e.,  the central 30$\arcsec$ surface densities  
after subtraction of the mean surrounding surface densities. 
For the IRAC color selection, the cumulative overdensity linearly increases with
increasing index of the 3C sources except for a small bump at
50\,$<$\,X\,$<$\,55, 
i.e. at  $z \sim 1.7$. 
The  overall  linear increase does not indicate a
significant redshift dependency except at $z \sim 1.7$, where the rest frame 1.6\,$\mu$m SED bump
shifts into the IRAC2 band, both increasing the 4.5\,$\mu$m flux and making 
the IRAC1/2 color particularly red
and thus the detection of $z \sim
1.7$ candidate cluster member galaxies most efficient, even in case of
IRAC1/2 color errors.  
On the other hand, for the PSO--IRAC  selection, the 
cumulative overdensity 
shows a change in slope, first being steep until X=35, i.e., $z \sim 1.4$,
and then at $z>1.4$ the increase becomes on average shallower until
X=61 ($z\sim2.5$) but again with a small bump at $z \sim 1.7$. 
(This bump is possibly due to increased 4.5\,$\mu$m flux as mentioned above).  
Comparison with 
Figure~\ref{fig_cumulative_densities} shows that the change of the
cumulative overdensity is mainly caused by the central surface density
rather than by the surrounding density.
This argues for a  
$z$-dependent decline of the measured overdensities
above  $z\sim 1.4$; we discuss that further below.

\subsection{Discussion}
\label{sec:discussion}

Both IRAC1/2 and 
PSO--IRAC  color criteria reveal, on average, a central overdensity of candidate cluster member
galaxies around the high-$z$ 3C sources in all redshift bins. 
This result is consistent with expectations from other high-$z$ AGN
samples.
Notably, the overdensities are better revealed by the
PSO--IRAC color selection which 
selects about a factor 1.4 more candidate
cluster member galaxies
than the IRAC color; this is evident from the
numbers listed below the histograms 
(Figs.~\ref{fig_single_radial_densities} and
~\ref{fig_single_radial_densities_without_overdensity}) and from the cumulative
density distributions (Fig.~\ref{fig_cumulative_densities}).
Evidently for our 3C data the IRAC color
selection not only rejects more sources but 
also some candidate cluster member galaxies. 
On the other hand,  the PSO$-$IRAC color selection rejects
fewer sources but appears to keep more true candidate cluster member
galaxies. To explain this, we
suggest that, besides the influence
of photometric color 
uncertainties mentioned above, 
also the type of rejected galaxies plays a
role: in case of color errors the IRAC1/2 color cut affects primarily
elliptical galaxies (lying close to the color cut in 
Fig.~\ref{fig_template_color_vs_redshift}~left), 
while the PSO--IRAC color cut affects primarily blue SF
galaxies (Fig.~\ref{fig_template_color_vs_redshift}~right). 
If 
cluster galaxies 
in our IRAC flux
limited sample are predominantly ellipticals, 
then  PSO--IRAC color errors will have little effect on the
PSO--IRAC 
selection, but IRAC color errors may have a large effect.
Therefore, we suggest that 
most of the actual cluster member galaxies detected by IRAC are
passive ellipticals rather than blue SF galaxies.

Further support for the passive nature of the cluster member galaxies
comes from the following check: we repeated the overdensity analysis
using a strong IRAC color cut  $[3.6] - [4.5] > -0.1$, i.e., 0.2 mag
redder than for the "soft"  
IRAC color cut  $[3.6] - [4.5] > -0.3$.
The strong IRAC color cut should remove
 ellipticals at $z \lesssim 1.5$
(Fig.~\ref{fig_template_color_vs_redshift}, left).
For the strong IRAC color cut, in fact, 
the cumulative central and surrounding surface densities perfectly
agree at  $1<z<1.4$ (Fig.~\ref{fig_cumulative_densities}). 
This is in contrast to the results for the soft  
IRAC color cut  $[3.6] - [4.5] > -0.3$ used before. 
This strongly suggests that the cluster member galaxies at $1<z<1.6$ are
passive ellipticals. 

If 
most IRAC-detected cluster galaxies are in fact passive, 
this might also explain why two
refinements of the IRAC1/2 color cut to select candidate 
cluster member galaxies failed to
yield more significant overdensities. These refinements are firstly
the combination of IRAC1/2 and PSO--IRAC color cuts and secondly
a redshift--dependent IRAC1/2 color cut
($-0.4+0.1z < [3.6] - [4.5] < -0.1+0.2z $) 
indicated 
in Figure~\ref{fig_template_color_vs_redshift}. 
In both cases, the additional constraints may better reject foreground
or surrounding galaxies but cannot recover the lost  
elliptical candidate cluster member galaxies.

Both the IRAC1/2 and PSO--IRAC color
cuts reveal a central 30$\arcsec$ overdensity for about half of the 3C sources
(33/61 for IRAC1/2 cut, 36/61 for PSO-IRAC cut, Table~\ref{table_overdensities}).  
An open issue is
whether for the remaining 3C sources  
overdensity is absent or escaped detection by our methods. 
If there were two classes, with and without overdensity,
one would expect that the overdensity
distribution is bimodal, but this is not seen in
Figure~\ref{fig_density_histogram}. 
The failure to see a bimodal distribution could be due to limited 
sensitivity of our methods and the low number
statistics.
Alternatively, an overdensity could be present in each 3C source but less concentrated (e.g., for  
proto-clusters), or there is a real spread in overdensities around the
3C sources as indicated by our visual inspection of the IRAC maps
(see also the negative overdensity for 3C\,298 in
Fig.~\ref{fig_single_radial_densities_without_overdensity}).

Clues on the limited sensitivity of our methods may come from  
the individual case of  3C\,270.1 at $z=1.5$, where the PSO-IRAC color
cut failed
to reveal a central (30$\arcsec$) overdensity,
the IRAC cut yields a marginal overdensity, and both cuts find an extended overdensity in the radial
bin at 30$\arcsec$ and 50$\arcsec$
(Fig.~\ref{fig_single_radial_densities_without_overdensity}, 
Table~\ref{table_overdensities}).
Both the elaborated IRAC environment study of 3C\,270.1
with supplementary deep $z$ and $y$ band imaging ($\sim$26 AB mag) and
photometric redshift determination from SED fitting of all sources in
the field 
\citep{Haas09} and the IRAC1/2 counts-in-cells analysis of the 3C\,270.1
field 
\citep{Galametz12} show an overdensity
within 50$\arcsec$ and 60$\arcsec$  radius, respectively. 
This suggests that both the IRAC1/2 and PSO--IRAC color
cuts are successful in revealing average overdensities but not
sensitive enough to detect overdensities for each individual 3C
source.  
Therefore, future investigations with improved methods and/or 
deeper data are needed.

For the PSO--IRAC selection of candidate cluster member galaxies, 
the cumulative central overdensity shows a redshift trend: 
The overdensity is largest at redshift $1.0 < z < 1.4$ and
declines at $z > 1.4$ (Fig.~\ref{fig_cumulative_densities_2}).
Such a decline appears consistent with expectations from cosmic
evolution models, e.g., passive galaxy evolution, 
as discussed for the CARLA sample at $1.3<z<3.2$ by \citet{Wylezalek14}. 
For our 3C data, 
however, 
the redshift dependent decline of the overdensity is not directly 
evident for the IRAC1/2 selection; 
the reason could be that as mentioned above in case of color errors the IRAC1/2 cut may reject elliptical cluster member galaxies at redshift $1.0 < z < 1.4$.

We examined the overdensities of the 27 quasars and 34 radio
galaxies separately (Table~\ref{tab_sample} lists the types Q and G). 
For quasars we find an overdensity with values for mean and uncertainty of
the mean of 
2.1 $\pm$ 0.6 (IRAC1/2) and 3.5 $\pm$ 0.7 (PSO--IRAC).
For radio galaxies the corresponding values are 
2.3 $\pm$ 0.7 (IRAC1/2) and 5.0 $\pm$ 1.0 (PSO--IRAC).
Thus, for both selection criteria, the mean overdensities 
of quasars and radio
galaxies agree within the uncertainties, 
consistent with the unified model \citep{Barthel89}.

Our derivation of the central
overdensities made the assumption that the 3C source is
located in the 
center of a potential cluster or proto-cluster. 
While  this
assumption reveals the  overdensities, 
visual inspection of the sky position of the candidate 
cluster members of the individual 3C sources indicates  
that  independent of redshift  in some cases the 3C 
source may be located at the cluster border. 
A future analysis of the individual 3C environments using density
maps with Voronoi cells
may provide further clues on the location of the 3C source within its 
cluster or proto-cluster. 

\section{Summary and outlook } 
\label{sec:summary}

We explored the environment of 
the complete sample of 64 high-redshift 3C radio galaxies and quasars 
as seen by the {\it Spitzer Space Telescope}, supplemented by 
Pan-STARRS (PSO).

IRAC color selection $[3.6] - [4.5] > -0.3$ AB mag and 
PSO--IRAC color selection $i - [4.5] > 0.5$ AB mag (plus all $i$-band
non-detections)
both reveal an average 
overdensity of galaxies within 250\,kpc projected distance from $z>1$ luminous
radio sources.
There is a clear field-to-field scatter in the overdensities of individual
3C environments; about half of the 3C sources show a significant (3-$\sigma$)
overdensity, and the other half do not. 
The origin of this difference has still to be explored.

The overdensities disappear at $1<z<1.4$ for a strong IRAC1/2 color
selection  $[3.6] - [4.5] > -0.1$ which removes passive elliptical galaxies at
that redshift but not star-forming galaxies. Therefore we conclude that
the cluster members are mainly passive elliptical galaxies.

The overdensities are more pronounced for the PSO--IRAC color
selection than for the IRAC color selection. 
The IRAC color selection rejects about 40\% more sources, but it appears to
reject also candidate cluster member galaxies. 
While primarily elliptical galaxies lie close to the IRAC color cuts,
the PSO--IRAC color cut primarily affects blue star forming
galaxies.
The cuts are quite sensitive to IRAC color errors but relatively
robust against PSO--IRAC color errors.
We suggest that the combination of IRAC color errors and the
prevalence of elliptical galaxies among the candidate cluster member
galaxies explains why the IRAC color leads to
lower overdensities than the 
PSO--IRAC color. 
Furthermore, the PSO--IRAC data indicate a decline of the average 
central overdensities with increasing redshift beyond $z \sim 1.4$,
consistent with galaxy evolution models.

Future investigations may provide further clues on the diversity of
the overdensities and the nature of the candidate cluster member galaxies. 
The {\it Herschel} far-infrared observations hold the key for dusty
starburst galaxies in the high-$z$ 3C environments (Westhues et al.\ in
prep.).
In addition, new {\it Hubble Space Telescope} images of a
sub-sample of high-$z$ 3Cs
with about 2$\arcmin$ FoV have been obtained \citep{Hilbert16}. 
The {\it Spitzer}$-$PSO study here finds that 
the overdensities occur within about 30$\arcsec$ radius
around the 3C sources. Therefore
the {\it HST} images might be large enough for a deep {\it HST-Spitzer}
optical-infrared study  of the overdensities (Ghaffari et
al.\  in prep.). 
Also, the analysis of {\it Chandra} X-ray data, where available for the 3C
sources, may reveal the extended 
hot X-ray gas characteristic of galaxy clusters (Wilkes et al.\ in prep).

~\\
{\it Acknowledgement: } 
This research has made use of the NASA/IPAC Extragalactic Database
(NED) and of the NASA/ IPAC Infrared Science Archive, which is
operated by the Jet Propulsion Laboratory, California 
Institute of Technology, under contract with the National Aeronautics
and Space Administration.  
This work is based in part on observations made with the {\it Spitzer
Space Telescope}, which was operated by the Jet Propulsion Laboratory
(JPL), Caltech under a contract with NASA. 
Some of the data presented in this paper were obtained from the
Mikulski Archive for Space Telescopes (MAST). STScI is operated by the
Association of Universities for Research in Astronomy, Inc., under
NASA contract NAS5-26555. Support for MAST for non-HST data is
provided by the NASA Office of Space Science via grant NNX09AF08G and
by other grants and contracts. 
The Pan-STARRS1 Surveys (PS1) and the PS1 public science archive have
been made possible through contributions by the Institute for Astronomy,
the University of Hawaii, the Pan-STARRS Project Office, the Max-Planck
Society and its participating institutes, the Max Planck Institute for
Astronomy, Heidelberg and the Max Planck Institute for Extraterrestrial
Physics, Garching, The Johns Hopkins University, Durham University, the
University of Edinburgh, the Queen's University Belfast, the
Harvard-Smithsonian Center for Astrophysics, the Las Cumbres Observatory
Global Telescope Network Incorporated, the National Central University
of Taiwan, the Space Telescope Science Institute, the National
Aeronautics and Space Administration under Grant No.\ NNX08AR22G issued
through the Planetary Science Division of the NASA Science Mission
Directorate, the National Science Foundation Grant No.\ AST-1238877, the
University of Maryland, Eotvos Lorand University (ELTE), the Los Alamos
National Laboratory, and the Gordon and Betty Moore Foundation.

\begin{table*}
\renewcommand{\thetable}{\arabic{table}}
\centering
\footnotesize
\caption{Number of candidate cluster member galaxies per radial bin
  for the color selections: IRAC1/2 $[3.6] - [4.5] > -0.3$ (columns N1-N6),  
  PSO--IRAC (columns M1-M6), strong IRAC1/2 $[3.6] - [4.5] > -0.1$ (columns K1-K6). 
  The six bin radii are 30, 50, 70, 90, 105, 120$\arcsec$. The six bin
  areas are 0.785, 1.396, 2.094, 2.793, 2.553, 2.945  square arcmin.
} 
\label{table_radial_densities}
\begin{tabular}{l|rrrrrr|rrrrrr|rrrrrr}
\hline
Name       & N1   & N2 & N3 & N4 & N5 & N6 &  M1 & M2 & M3 & M4 & M5 & M6   & K1 & K2 & K3 & K4 & K5 & K6  \\
\hline 
     3C\,002 &    9 &     7 &   14 &  15 &  16 &  14  &   12 &  11 &  17 &  19 &  21 &  18  &    4 &   5 &  11 &  10 &  12 &   8 \\
     3C\,009 &   11 &    18 &   26 &  19 &  34 &  27  &   12 &  22 &  32 &  24 &  39 &  36  &    9 &  14 &  23 &  16 &  26 &  20 \\
     3C\,013 &    8 &     6 &   11 &  23 &  18 &  13  &    9 &   8 &  11 &  30 &  21 &  19  &    5 &   3 &   7 &  12 &   8 &   6 \\
     3C\,014 &    4 &    11 &   11 &  12 &  11 &  14  &    4 &  14 &  22 &  17 &  21 &  16  &    3 &   9 &   9 &   6 &   9 &   9 \\
     3C\,036 &    8 &    12 &   13 &  21 &  14 &  17  &    9 &  12 &  13 &  25 &  24 &  22  &    5 &   7 &   8 &  11 &   5 &  10 \\
     3C\,043 &    8 &     6 &    6 &   6 &  11 &  14  &    8 &   5 &   9 &  13 &  12 &  20  &    6 &   6 &   5 &   4 &   9 &  13 \\
     3C\,065 &    6 &     9 &   12 &  15 &  11 &  22  &   11 &  13 &  18 &  20 &  13 &  24  &    5 &   5 &   8 &   9 &   4 &  13 \\
   3C\,068.1 &    4 &    19 &   12 &  15 &  16 &  19  &    6 &  23 &  20 &  22 &  21 &  27  &    1 &  11 &   4 &   7 &   6 &  12 \\
   3C\,068.2 &    7 &     6 &    9 &  11 &  13 &  15  &   13 &   9 &  17 &  19 &  24 &  26  &    5 &   1 &   7 &   5 &   7 &   9 \\
     3C\,119 &    5 &    11 &   10 &  12 &  15 &  12  &    9 &  11 &  14 &  21 &  25 &  15  &    2 &   6 &   6 &   4 &   7 &   4 \\
     3C\,124 &   11 &     5 &   19 &  17 &  23 &  22  &   15 &  12 &  19 &  26 &  27 &  30  &    3 &   4 &  13 &  11 &  12 &  13 \\
     3C\,173 &    2 &     7 &   12 &  15 &  13 &  20  &    6 &  11 &  14 &  22 &  14 &  22  &    2 &   3 &   7 &   8 &   9 &  10 \\
     3C\,181 &    8 &     9 &   12 &  16 &  22 &  11  &    9 &  12 &  20 &  22 &  27 &  18  &    6 &   5 &   8 &   9 &  11 &   5 \\
     3C\,186 &    3 &    10 &    8 &   9 &   9 &  13  &    5 &  14 &  12 &  11 &  11 &  16  &    2 &   6 &   4 &   6 &   8 &   5 \\
     3C\,190 &    7 &     3 &   21 &  18 &  17 &  20  &   10 &   7 &  26 &  23 &  22 &  24  &    4 &   1 &  14 &  12 &  13 &  10 \\
     3C\,191 &    6 &     5 &   13 &  13 &  17 &  20  &    9 &   7 &  12 &  16 &  22 &  21  &    4 &   5 &   7 &   5 &  12 &  13 \\
     3C\,194 &    4 &    11 &   15 &  25 &  21 &  22  &    9 &  12 &  16 &  25 &  24 &  25  &    2 &   7 &   7 &  14 &  12 &  14 \\
     3C\,204 &   11 &    12 &   23 &  16 &  27 &  25  &   16 &  16 &  31 &  21 &  34 &  25  &    1 &   5 &  11 &  11 &  13 &  14 \\
     3C\,205 &    5 &    11 &   14 &  13 &  15 &  18  &   10 &  16 &  19 &  22 &  21 &  17  &    4 &   7 &   9 &  12 &   6 &  11 \\
   3C\,208.1 &    7 &    14 &   10 &  29 &  15 &  24  &    9 &  14 &  18 &  36 &  19 &  27  &    6 &   9 &   7 &  22 &  10 &  16 \\
     3C\,208 &   12 &    10 &   10 &  17 &  17 &  21  &   25 &  12 &  10 &  15 &  17 &  26  &    7 &   5 &   5 &  11 &  11 &  15 \\
     3C\,210 &    9 &    12 &   28 &  36 &  27 &  27  &   26 &  16 &  41 &  46 &  31 &  41  &    2 &   5 &  13 &  20 &  17 &  13 \\
     3C\,212 &    9 &    15 &    7 &  19 &  13 &  14  &   13 &  17 &   9 &  23 &  19 &  22  &    3 &  10 &   5 &  13 &   8 &   8 \\
   3C\,220.2 &    8 &    13 &   15 &  14 &  13 &  18  &   11 &  14 &  17 &  14 &  14 &  25  &    5 &   7 &  11 &  11 &   9 &  15 \\
     3C\,222 &    7 &     4 &    5 &  17 &  12 &  15  &   12 &   6 &   8 &  22 &  15 &  15  &    5 &   4 &   2 &  13 &   7 &  12 \\
    3C\,225A &    3 &     7 &    5 &  25 &  22 &  14  &    3 &   7 &   6 &  27 &  24 &  18  &    2 &   3 &   4 &  12 &  10 &  11 \\
     3C\,230 &    5 &    11 &    7 &  12 &  10 &  14  &    7 &  11 &   9 &  17 &  19 &  18  &    1 &   6 &   4 &   6 &   5 &  10 \\
     3C\,238 &    9 &     9 &    5 &  13 &  15 &  12  &   11 &  11 &   5 &  19 &  17 &  19  &    4 &   4 &   2 &   6 &  11 &   9 \\
     3C\,241 &   11 &    13 &   13 &  25 &  18 &  23  &   12 &  14 &  15 &  32 &  21 &  24  &    8 &   8 &  10 &  14 &  10 &  15 \\
     3C\,245 &    8 &    11 &   18 &  25 &  21 &  19  &   13 &  11 &  24 &  32 &  28 &  33  &    3 &   7 &  12 &  19 &  13 &  11 \\
     3C\,249 &    9 &     4 &    5 &  16 &  15 &  16  &   10 &   6 &  10 &  22 &  19 &  15  &    5 &   3 &   3 &  12 &   7 &  12 \\
     3C\,250 &    9 &     8 &   13 &  15 &   7 &  20  &   11 &  10 &  15 &  17 &  13 &  22  &    4 &   7 &   6 &  10 &   5 &  13 \\
     3C\,252 &    2 &    12 &   14 &  18 &  16 &  26  &    8 &  15 &  19 &  21 &  20 &  32  &    0 &   7 &  10 &   8 &  13 &  20 \\
     3C\,255 &    5 &    14 &   14 &  19 &  23 &  34  &   11 &  19 &  15 &  27 &  26 &  43  &    2 &  10 &   8 &  10 &  16 &  26 \\
     3C\,256 &    5 &    21 &   13 &  25 &  28 &  29  &   10 &  24 &  21 &  35 &  37 &  46  &    2 &  13 &   7 &  13 &  17 &  12 \\
     3C\,257 &    4 &     1 &    9 &  12 &  15 &  12  &    5 &   6 &   9 &  15 &  16 &  20  &    3 &   1 &   7 &   8 &  10 &   9 \\
     3C\,266 &    2 &     6 &   15 &  13 &  18 &  12  &    4 &   8 &  17 &  21 &  24 &  26  &    1 &   4 &  12 &   8 &  10 &   8 \\
     3C\,267 &   10 &    14 &   20 &  31 &  24 &  17  &   14 &  15 &  25 &  37 &  29 &  25  &    5 &   9 &  14 &  26 &  18 &  14 \\
   3C\,268.4 &    4 &     9 &    9 &  10 &  11 &  15  &    7 &   9 &  16 &  11 &  19 &  25  &    2 &   6 &   6 &   7 &   3 &   9 \\
   3C\,270.1 &    7 &    16 &   17 &  17 &  15 &  18  &    8 &  20 &  21 &  26 &  20 &  29  &    4 &  13 &   9 &  11 &  10 &  13 \\
   3C\,280.1 &    8 &    13 &   20 &  19 &  17 &  15  &    9 &  17 &  21 &  21 &  21 &  20  &    7 &   8 &  14 &  14 &  12 &   8 \\
     3C\,287 &    8 &    11 &   14 &  15 &  24 &  24  &    9 &  13 &  18 &  25 &  28 &  31  &    4 &   8 &   9 &  14 &  15 &  13 \\
     3C\,297 &    4 &    11 &   14 &  14 &  13 &  24  &    4 &  12 &  15 &  18 &  16 &  30  &    3 &   8 &  13 &  12 &  10 &  14 \\
     3C\,298 &    3 &     6 &   16 &  17 &  19 &  14  &    5 &  10 &  17 &  19 &  27 &  16  &    3 &   4 &   7 &   6 &  12 &   7 \\
   3C\,300.1 &    6 &    10 &   20 &  20 &  23 &  33  &   14 &  19 &  26 &  22 &  25 &  39  &    2 &   5 &  10 &  10 &  13 &  26 \\
   3C\,305.1 &    5 &     7 &   17 &  14 &  16 &  28  &    6 &  12 &  24 &  19 &  21 &  35  &    2 &   4 &  11 &  13 &   8 &  15 \\
     3C\,318 &    6 &     5 &   16 &  18 &  16 &  17  &    5 &   9 &  18 &  21 &  17 &  20  &    5 &   3 &  14 &  14 &  13 &  10 \\
     3C\,322 &   10 &    21 &   16 &  15 &  11 &  20  &   12 &  23 &  19 &  19 &  17 &  23  &    7 &  13 &  11 &   8 &   8 &  15 \\
     3C\,324 &   12 &    13 &   19 &  32 &  32 &  33  &   17 &  18 &  22 &  34 &  42 &  38  &    3 &   5 &  12 &  24 &  18 &  18 \\
     3C\,325 &   10 &     6 &   17 &  22 &  18 &  19  &   14 &  12 &  22 &  38 &  29 &  26  &    2 &   5 &   7 &  11 &  11 &  13 \\
   3C\,326.1 &   11 &     8 &   24 &  24 &  16 &  18  &   13 &  11 &  26 &  31 &  16 &  21  &    8 &   4 &  19 &  13 &  10 &  10 \\
     3C\,356 &    3 &     8 &   18 &   9 &  15 &  25  &    7 &   6 &  21 &  10 &  16 &  22  &    1 &   3 &  12 &   3 &  13 &  16 \\
     3C\,368 &    4 &     8 &    3 &  14 &   3 &  20  &    4 &   8 &   7 &  12 &   7 &  18  &    4 &   2 &   1 &   6 &   1 &  13 \\
     3C\,418 &   10 &     5 &    3 &   8 &   9 &  10  &   14 &   9 &  10 &  22 &  16 &  22  &    4 &   0 &   0 &   3 &   2 &   4 \\
     3C\,432 &    5 &     9 &   10 &  16 &  13 &  13  &    5 &   9 &  11 &  21 &  18 &  15  &    1 &   5 &   6 &   6 &   5 &  11 \\
     3C\,437 &    4 &    12 &   11 &  13 &  13 &  16  &    5 &  15 &  15 &  14 &  17 &  20  &    3 &  10 &   9 &   9 &   7 &   9 \\
   3C\,454.1 &   10 &     6 &    9 &  11 &  18 &   9  &    9 &   8 &  12 &  14 &  21 &  12  &    3 &   4 &   6 &   9 &  10 &   4 \\
     3C\,454 &    5 &     4 &   12 &  11 &  13 &  10  &    6 &   9 &  13 &  15 &  18 &  17  &    3 &   1 &   8 &   9 &  11 &   8 \\
   3C\,469.1 &    7 &     9 &   20 &  23 &  20 &  27  &   16 &  17 &  21 &  41 &  25 &  36  &    3 &   6 &  14 &  15 &  13 &  18 \\
   4C\,13.66 &    8 &     5 &   13 &  18 &  22 &   8  &    7 &   8 &  19 &  23 &  19 &  20  &    4 &   4 &   7 &   8 &   7 &   2 \\
   4C\,16.49 &    4 &    12 &   12 &  30 &  28 &  30  &   13 &  13 &  17 &  31 &  31 &  39  &    3 &   5 &   7 &  12 &  13 &  19 \\
\hline
\end{tabular}
\end{table*}

\begin{table*}
\renewcommand{\thetable}{\arabic{table}}
\centering
\footnotesize
\caption{Central and average surrounding surface densities per square arcmin. 
Columns 2--6 for the IRAC1/2 $[3.6] - [4.5] > -0.3$ cut, 
cols. 7--11 for the PSO--IRAC color selection,
cols. 12--16 for the strong IRAC1/2 $[3.6] - [4.5] > -0.1$ cut.
Columns 2, 7, 12 in the central bin of 30$\arcsec$ radius.
Columns 3, 8, 13 in the 30$\arcsec$--50$\arcsec$ annulus.
Columns Surr and Err give mean and error of the mean of the 
surrounding surface densities, calculated from the 50$\arcsec$--120$\arcsec$ annuli.
Columns 6, 11, 12 indicate a significant central 30$\arcsec$ overdensity above surrounding + 3$\sigma$ ($+$), or not ($-$).
} 
\label{table_overdensities}
\begin{tabular}{l|rrrrr|rrrrr|rrrrr}
\hline
Name       &  30$\arcsec$   & 50$\arcsec$ &  Surr &  Err &  OD  &  30$\arcsec$   &  50$\arcsec$ &  Surr &  Err &  OD & 30$\arcsec$   & 50$\arcsec$ &  Surr &  Err &  OD  \\
1       & 2  & 3  & 4 &  5 & 6  & 7  & 8& 9 & 10 & 11 & 12  & 13  & 14 &  15 & 16 \\
\hline     
     3C\,002 &  11.46 &   5.01 &   5.77 &   0.44 & +  &  15.28 &   7.88 &   7.31 &   0.52 & +  &   5.09 &   3.58 &   4.06 &   0.57 & -   \\
     3C\,009 &  14.01 &  12.89 &  10.43 &   1.50 & -  &  15.28 &  15.76 &  12.76 &   1.60 & -  &  11.46 &  10.03 &   8.42 &   1.28 & -   \\
     3C\,013 &  10.19 &   4.30 &   6.24 &   0.86 & +  &  11.46 &   5.73 &   7.67 &   1.19 & +  &   6.37 &   2.15 &   3.20 &   0.46 & +   \\
     3C\,014 &   5.09 &   7.88 &   4.65 &   0.23 & -  &   5.09 &  10.03 &   7.44 &   1.05 & -  &   3.82 &   6.45 &   3.26 &   0.45 & -   \\
     3C\,036 &  10.19 &   8.59 &   6.25 &   0.45 & +  &  11.46 &   8.59 &   7.81 &   0.57 & +  &   6.37 &   5.01 &   3.28 &   0.46 & +   \\
     3C\,043 &  10.19 &   4.30 &   3.52 &   0.61 & +  &  10.19 &   3.58 &   4.99 &   0.63 & +  &   7.64 &   4.30 &   2.94 &   0.65 & +   \\
     3C\,065 &   7.64 &   6.45 &   5.72 &   0.66 & -  &  14.01 &   9.31 &   7.07 &   0.84 & +  &   6.37 &   3.58 &   3.26 &   0.61 & +   \\
   3C\,068.1 &   5.09 &  13.61 &   5.96 &   0.25 & -  &   7.64 &  16.47 &   8.51 &   0.50 & -  &   1.27 &   7.88 &   2.71 &   0.47 & -   \\
   3C\,068.2 &   8.91 &   4.30 &   4.51 &   0.25 & +  &  15.28 &   5.01 &   8.19 &   0.50 & +  &   6.37 &   0.72 &   2.73 &   0.34 & +   \\
     3C\,119 &   6.37 &   7.88 &   4.47 &   0.52 & +  &  11.46 &   6.45 &   5.68 &   1.06 & +  &   2.55 &   4.30 &   2.10 &   0.41 & -   \\
     3C\,124 &  14.01 &   3.58 &   7.81 &   0.67 & +  &  17.83 &   8.59 &   9.43 &   0.28 & +  &   3.82 &   2.86 &   4.82 &   0.49 & -   \\
     3C\,173 &   2.55 &   4.30 &   5.66 &   0.41 & -  &   7.64 &   7.88 &   6.70 &   0.44 & -  &   2.55 &   2.15 &   3.28 &   0.14 & -   \\
     3C\,181 &  10.19 &   6.45 &   5.95 &   1.01 & +  &  11.46 &   7.88 &   8.53 &   0.98 & -  &   7.64 &   3.58 &   3.26 &   0.57 & +   \\
     3C\,186 &   3.82 &   7.16 &   3.54 &   0.19 & -  &   6.37 &  10.03 &   4.85 &   0.43 & +  &   2.55 &   4.30 &   2.22 &   0.32 & -   \\
     3C\,190 &   8.91 &   2.15 &   7.48 &   0.85 & -  &  12.73 &   4.30 &   9.35 &   1.02 & +  &   5.09 &   0.72 &   4.87 &   0.70 & -   \\
     3C\,191 &   7.64 &   3.58 &   6.08 &   0.49 & +  &  11.46 &   5.01 &   6.77 &   0.73 & +  &   5.09 &   3.58 &   3.56 &   0.66 & -   \\
     3C\,194 &   5.09 &   7.88 &   7.95 &   0.40 & -  &  11.46 &   8.59 &   8.62 &   0.38 & +  &   2.55 &   5.01 &   4.45 &   0.38 & -   \\
     3C\,204 &  14.01 &   7.88 &   8.86 &   1.22 & +  &  20.37 &  11.46 &  10.94 &   1.85 & +  &   1.27 &   3.58 &   4.76 &   0.29 & -   \\
     3C\,205 &   6.37 &   7.88 &   5.83 &   0.43 & -  &  12.73 &  11.46 &   7.57 &   0.77 & +  &   5.09 &   5.01 &   3.67 &   0.46 & +   \\
     3C\,208 &  15.28 &   7.16 &   6.16 &   0.51 & +  &  31.83 &   8.59 &   6.19 &   0.97 & +  &   8.91 &   3.58 &   3.93 &   0.57 & +   \\
   3C\,208.1 &   8.91 &  10.03 &   7.21 &   1.23 & -  &  11.46 &   9.31 &   9.35 &   1.21 & -  &   7.64 &   6.45 &   5.14 &   1.01 & -   \\
     3C\,210 &  11.46 &   8.59 &  11.59 &   1.03 & -  &  33.10 &  11.46 &  15.22 &   1.57 & +  &   2.55 &   3.58 &   6.11 &   0.60 & -   \\
     3C\,212 &  11.46 &  10.74 &   4.88 &   0.81 & +  &  15.28 &  12.18 &   6.74 &   0.99 & +  &   3.82 &   7.16 &   3.22 &   0.50 & -   \\
   3C\,220.2 &  10.19 &   8.59 &   5.84 &   0.51 & +  &  14.01 &   9.31 &   6.78 &   0.89 & +  &   6.37 &   5.01 &   4.45 &   0.43 & +   \\
     3C\,222 &   8.91 &   2.86 &   4.57 &   0.78 & +  &  15.28 &   4.30 &   5.67 &   0.85 & +  &   6.37 &   2.86 &   3.11 &   0.82 & +   \\
    3C\,225A &   2.55 &   5.01 &   6.18 &   1.58 & -  &   2.55 &   5.01 &   7.01 &   1.60 & -  &   2.55 &   2.15 &   3.46 &   0.53 & -   \\
     3C\,230 &   5.09 &   7.88 &   4.08 &   0.30 & +  &   7.64 &   7.88 &   5.90 &   0.65 & -  &   1.27 &   4.30 &   2.35 &   0.35 & -   \\
     3C\,238 &  11.46 &   6.45 &   4.25 &   0.73 & +  &  14.01 &   7.88 &   5.58 &   1.07 & +  &   5.09 &   2.86 &   2.62 &   0.71 & +   \\
     3C\,241 &  14.01 &   9.31 &   7.51 &   0.58 & +  &  15.28 &  10.03 &   8.75 &   0.94 & +  &  10.19 &   5.73 &   4.70 &   0.27 & +   \\
     3C\,245 &  10.19 &   7.88 &   7.97 &   0.51 & +  &  16.55 &   7.88 &  11.18 &   0.10 & +  &   3.82 &   5.01 &   5.34 &   0.64 & -   \\
     3C\,249 &  11.46 &   2.86 &   4.86 &   0.83 & +  &  12.73 &   4.30 &   6.21 &   0.74 & +  &   6.37 &   2.15 &   3.14 &   0.66 & +   \\
     3C\,250 &  11.46 &   5.73 &   5.28 &   0.89 & +  &  14.01 &   7.16 &   6.45 &   0.54 & +  &   5.09 &   5.01 &   3.20 &   0.52 & +   \\
     3C\,252 &   2.55 &   8.59 &   7.06 &   0.60 & -  &  10.19 &  10.74 &   8.82 &   0.76 & -  &   0.00 &   5.01 &   4.88 &   0.80 & -   \\
     3C\,255 &   6.37 &  10.03 &   8.43 &   1.07 & -  &  12.73 &  13.61 &  10.31 &   1.55 & -  &   2.55 &   7.16 &   5.62 &   1.23 & -   \\
     3C\,256 &   6.37 &  15.04 &   8.99 &   1.02 & -  &  12.73 &  17.19 &  13.17 &   1.16 & -  &   2.55 &   9.31 &   4.68 &   0.71 & -   \\
     3C\,257 &   5.09 &   0.72 &   4.64 &   0.42 & -  &   6.37 &   5.01 &   5.60 &   0.49 & -  &   3.82 &   0.72 &   3.30 &   0.23 & -   \\
     3C\,266 &   2.55 &   4.30 &   5.74 &   0.80 & -  &   5.09 &   5.73 &   8.20 &   0.64 & -  &   1.27 &   2.86 &   3.81 &   0.69 & -   \\
     3C\,267 &  12.73 &  10.03 &   9.13 &   1.11 & +  &  17.83 &  10.74 &  11.43 &   0.99 & +  &   6.37 &   6.45 &   6.95 &   0.93 & -   \\
   3C\,268.4 &   5.09 &   6.45 &   4.32 &   0.31 & -  &   8.91 &   6.45 &   6.88 &   1.01 & -  &   2.55 &   4.30 &   2.40 &   0.42 & -   \\
   3C\,270.1 &   8.91 &  11.46 &   6.55 &   0.53 & +  &  10.19 &  14.32 &   9.35 &   0.40 & -  &   5.09 &   9.31 &   4.14 &   0.13 & +   \\
   3C\,280.1 &  10.19 &   9.31 &   7.03 &   0.93 & +  &  10.19 &  12.18 &   8.23 &   0.64 & +  &   8.91 &   5.73 &   4.78 &   0.81 & +   \\
     3C\,287 &  10.19 &   7.16 &   7.40 &   0.88 & +  &  11.46 &   9.31 &   9.77 &   0.63 & -  &   5.09 &   5.73 &   4.90 &   0.36 & -   \\
     3C\,297 &   5.09 &   7.88 &   6.23 &   0.75 & -  &   5.09 &   8.59 &   7.40 &   0.93 & -  &   3.82 &   5.73 &   4.79 &   0.50 & -   \\
     3C\,298 &   3.82 &   4.30 &   6.48 &   0.67 & -  &   6.37 &   7.16 &   7.73 &   1.10 & -  &   3.82 &   2.86 &   3.14 &   0.58 & -   \\
   3C\,300.1 &   7.64 &   6.45 &   9.01 &   0.84 & -  &  17.83 &  12.89 &  10.66 &   1.12 & +  &   2.55 &   3.58 &   5.57 &   1.13 & -   \\
   3C\,305.1 &   6.37 &   5.01 &   7.23 &   0.99 & -  &   7.64 &   8.59 &   9.38 &   1.22 & -  &   2.55 &   2.86 &   4.53 &   0.48 & -   \\
     3C\,318 &   7.64 &   3.58 &   6.53 &   0.40 & -  &   5.09 &   6.45 &   7.30 &   0.44 & -  &   6.37 &   2.15 &   5.05 &   0.67 & -   \\
     3C\,322 &  12.73 &  15.04 &   6.03 &   0.74 & +  &  15.28 &  16.47 &   7.59 &   0.56 & +  &   8.91 &   9.31 &   4.09 &   0.63 & +   \\
     3C\,324 &  15.28 &   9.31 &  11.07 &   0.73 & +  &  21.65 &  12.89 &  13.01 &   1.25 & +  &   3.82 &   3.58 &   6.87 &   0.64 & -   \\
     3C\,325 &  12.73 &   4.30 &   7.37 &   0.38 & +  &  17.83 &   8.59 &  10.89 &   0.91 & +  &   2.55 &   3.58 &   4.00 &   0.24 & -   \\
   3C\,326.1 &  14.01 &   5.73 &   8.01 &   1.30 & +  &  16.55 &   7.88 &   8.84 &   1.48 & +  &  10.19 &   2.86 &   5.26 &   1.30 & +   \\
     3C\,356 &   3.82 &   5.73 &   6.55 &   1.27 & -  &   8.91 &   4.30 &   6.84 &   1.34 & -  &   1.27 &   2.15 &   4.33 &   1.09 & -   \\
     3C\,368 &   5.09 &   5.73 &   3.60 &   1.38 & -  &   5.09 &   5.73 &   3.58 &   0.95 & -  &   5.09 &   1.43 &   1.86 &   0.94 & +   \\
     3C\,418 &   8.91 &   2.86 &   2.32 &   0.46 & +  &  10.19 &   2.15 &   3.54 &   0.18 & +  &   5.09 &   0.00 &   0.80 &   0.29 & +   \\
     3C\,432 &   6.37 &   6.45 &   5.00 &   0.28 & +  &   6.37 &   6.45 &   6.14 &   0.56 & -  &   1.27 &   3.58 &   2.68 &   0.40 & -   \\
     3C\,437 &   5.09 &   8.59 &   5.14 &   0.31 & -  &   6.37 &  10.74 &   6.31 &   0.47 & -  &   3.82 &   7.16 &   3.33 &   0.34 & -   \\
   3C\,454.1 &  11.46 &   4.30 &   4.49 &   0.77 & +  &  10.19 &   5.01 &   5.11 &   0.79 & +  &   3.82 &   2.86 &   2.84 &   0.54 & -   \\
     3C\,454 &   6.37 &   2.86 &   4.54 &   0.53 & +  &   7.64 &   5.01 &   6.02 &   0.39 & +  &   3.82 &   0.72 &   3.52 &   0.35 & -   \\
   3C\,469.1 &   8.91 &   6.45 &   8.70 &   0.40 & -  &  20.37 &  11.46 &  11.75 &   1.04 & +  &   3.82 &   4.30 &   5.82 &   0.36 & -   \\
   4C\,13.66 &  10.19 &   3.58 &   5.81 &   1.23 & +  &   8.91 &   5.01 &   7.23 &   0.58 & -  &   5.09 &   2.86 &   2.41 &   0.59 & +   \\
   4C\,16.49 &   5.09 &   8.59 &   9.19 &   1.32 & -  &  16.55 &   9.31 &  10.97 &   1.02 & +  &   3.82 &   3.58 &   4.80 &   0.66 & -   \\
   \hline
\end{tabular}
\end{table*}


\bibliographystyle{an}
\bibliography{biblio.bib}


\end{document}